\documentclass[11pt,authoryear]{elsarticle}

\usepackage{url}
\usepackage{graphicx}
\usepackage[psamsfonts]{amssymb}
\usepackage{amsmath}
\usepackage{indentfirst}

\usepackage{booktabs}
\usepackage{tabularx}
\usepackage{multirow} 

\usepackage[margin=3cm]{geometry}

\usepackage{color}

\journal{Special Issue on Extreme Events}

\begin{document}
\begin{frontmatter}

\title{Power law scaling and ``Dragon-Kings'' \\ in distributions of intraday financial drawdowns}

\author[1]{Vladimir Filimonov}
\ead{vfilimonov@ethz.ch}

\author[1,2]{Didier Sornette}
\ead{dsornette@ethz.ch}

\address[1]{Dept. of Management, Technology and Economics, ETH Z\"{u}rich, Z\"{u}rich, Switzerland}
\address[2]{Swiss Finance Institute, c/o University of Geneva}

\date{\today}

\begin{abstract}
We investigate the distributions of $\epsilon$-drawdowns and $\epsilon$-drawups of 
the most liquid futures financial contracts of the world at time scales of $30$ seconds.
The $\epsilon$-drawdowns (resp. $\epsilon$-drawups) generalise the notion of runs
of negative (resp. positive) returns so as to capture the risks
to which investors are arguably the most concerned with.  
Similarly to the distribution of returns, we find that the distributions of 
$\epsilon$-drawdowns and $\epsilon$-drawups exhibit power law tails, albeit
with exponents significantly larger than those for the return distributions.
This paradoxical result can be attributed to (i) the existence of significant
transient dependence between returns and (ii) the presence of large outliers (dragon-kings) characterizing
the extreme tail of the drawdown/drawup distributions deviating from the power law.
The study of the tail dependence between the
sizes, speeds and durations of drawdown/drawup indicates a clear relationship between size and speed 
but none between size and duration. This implies that the most extreme 
drawdown/drawup tend to occur fast and are dominated by a few very large returns.
We discuss both the endogenous and exogenous origins of these extreme events.
\end{abstract}


\end{frontmatter}
\clearpage
%\maketitle

%\todo{\url{http://www.journals.elsevier.com/chaos-solitons-and-fractals/call-for-papers/special-issue-on-extreme-events/}}

%===============================================================================
\section{Introduction}

Traditionally, market risk is proxied by the distribution of asset log-returns $r_t^{(\Delta t)}$ on different scales $\Delta t$. Such distribution for most assets in various classes is well-known to have a power law tail $\mathrm{Pr}[r_t^{(\Delta t)} >x] \sim x^{-\mu}$
with the ``tail exponent'' $\mu$ in the range from 3 to 5 both for daily~\citep{deVries1994, Pagan1996, Gopikrishnan1998, Cont2001,Sornette2005} and intraday returns~\citep{Chakraborti2011_1}. Though  the second and possibly third order moments of the distribution exist, the traditional volatility measures based on variance of returns are not sufficient for quantifying the risk associated with extreme events. 

Much better metrics to capture systematic events are the so-called \emph{drawdowns} (and their complements, the \emph{drawups}), which are traditionally defined as a persistent decrease (respectively increase) in the price over consecutive $\Delta t$-time intervals \citep{JohansenSornette2001JofRisk}. In other words, a drawdown is the cumulative loss from the last maximum to the next minimum of the price, and a drawup is the the price change between a local minimum and the following maximum. By definition, drawups and drawdowns alternate: a drawdown follows a drawup and vice versa.

In contrast to simple returns, drawdowns are much more flexible measures of risk as they also capture the transient time-dependence of consecutive price changes. Drawdowns quantify the worst-case scenario of an investor buying at the local high and selling at the next minimum (similarly drawups quantifies the upside potential of buying at the lowest price and selling at the highest one). The duration of drawdowns is not fixed as well: some drawdowns can end in one drop of duration $\Delta t$, when others may last for tens to hundreds $\Delta t$'s. The distribution of drawdowns contains information that is quite different from the distribution of returns over a fixed time scale. In particular, a drawdown reflects a transient ``memory'' of the market by accounting for possible inter-dependence during series of losses~\citep{SornetteJohansen2000,JohansenSornette2001JofRisk}. During crashes, 
positive feedback mechanisms are activated so that previous losses lead to further selling, strengthening the downward trend, as for instance as a result of the implementation of so-called
portfolio insurance strategies \citep{Lelandrub}. The resulting drawdowns will capture
these transient positive feedbacks, much more than the returns or even the
two-point correlation functions between returns or between volatilities.
In contrast to autocorrelation measures, which quantify the average (or global) serial 
linear dependence between returns over a generally large chosen time period, 
drawdowns are local measures, i.e. they account for rather instantaneous 
dependences between returns that are specific to a given event. Statistically, drawdowns are 
related to the notion of ``runs'' that is often used in econometrics~\citep{Campbell1996}.

This paper presents an analysis of the statistical properties of intraday drawdowns and drawups. Our tests are performed on the most liquid Futures Contracts of the world. Our results are thus of general relevance
and are offered as novel ``stylized facts'' of the price dynamics.
We discuss and quantify the distribution of intra-day extreme events, and compare distributional characteristics of drawdowns with those of individual returns. In so doing, we discover that
the generally accepted description of the tail of the distribution of returns
by a power law distribution is incorrect: we find highly statistically significant 
upward deviations from the power law by the most extreme events. These
deviations are associated with well-known events, such as the ``flash-crash'' of May 6, 2010.
Statistical tests designed to detect such deviations 
confirm their high significance, implying that these events belong to a special class of 
so-called \emph{``Dragon-Kings''} \citep{Sornette2009,SorouillonDK12}: these events are generated with different amplifying mechanisms than the rest of the population. We show that some of these events can be attributed to an internal mutual-excitation between market participants, while others are pure response to external news.  As for extreme drawdowns, 
there are in principle two end-member generating mechanisms for them:
(i) one return in the run of losses is an extreme loss and, by itself alone, makes the drawdown
extreme; (ii) rare transient dependences between negative returns make some runs 
especially large. We  document that most of the extreme drawdowns are generated 
by the second mechanism, that is, by emerging spontaneous correlation patterns, rather than by 
the domination of one or a few extreme individual returns.

The paper is organized as follows. Section~\ref{sec:data} discusses the high-frequency data and cleaning procedures. Section~\ref{sec:dd} presents the detection method of the so-called \emph{$\epsilon$-drawdowns} that we use as a proxy of transient directional price movements. Section~\ref{sec:descript_stat} provides descriptive statistics of the detected events. Section~\ref{sec:distrib_dd} focuses on the properties of the distributions of drawdowns and quantify their tails as belonging essentially to a power law regime. In section~\ref{sec:dragonkings}, we present a generalised Dragon-King test (DK-test), derived and improved from 
\citep{PisarenkoSornette2012}, which allows us to quantify the statistical significance of 
the deviations of extreme drawdowns from the power law distribution calibrated
on the rest of the distribution. Section~\ref{sec:distrib_aggragated} describes the aggregated distributions 
over all tickers and validates that our findings hold both at individual and global levels. For this, we use the generalised DK-test as well as the parametric U-test also introduced by \cite{PisarenkoSornette2012}
and study their respective complementary merits. Section~\ref{sec:dependence} examines the interdependence of the speed and durations of extreme drawdowns with respect to their size. Section~\ref{sec:conclusion} concludes.

%===============================================================================
\section{The data}\label{sec:data}

We use tick data for the most actively traded Futures Contracts on the World Indices (see Table~\ref{tb:contracts}) from January 1, 2005 to December 30, 2011. For Futures on the OMX Stockholm 30 Index (OMXS), our dataset starts from September 1, 2005; and for Futures on Hong Kong indexes (HSI and HCEI), our datasets are limited to the period before April 1, 2011.  For Futures on the BOVESPA index (BOVESPA), we restrict our analysis to the period after January 1, 2009, ignoring the relatively inactive trading in 2005--2008.

Many of the contracts presented in Table \ref{tb:contracts} are traded almost continuously (e.g. E-mini S\&P 500 futures contracts are traded every business day from Monday to Friday with only two trading halts: from 16:15 to 17:00 CDT and from 15:15 to 15:30 CDT). Though it is being progressively changing~\citep{FilimonovSornette2013_apparent}, most of the daily volume is traded within so-called Regular Trading Hours (RTH, in case of E-mini contracts: 8:30--15:15 CDT). For Asian exchanges, the activity within Regular Trading Hours is also non-uniform. For the analysis, we have limited ourselves only to the part of RTH where the trading is the most active (in terms of volume), which we refer to as an Active Trading Hours (ATH) in Table~\ref{tb:contracts}.

\begin{table}[th!]
%\vspace{-2ex}
\caption{Description of the Futures Contracts used for analysis.}
\label{tb:contracts}
\begin{center}
\footnotesize
%\scriptsize
%\renewcommand{\arraystretch}{0.8}

% \multirow{9}{*}{Europe}
% \multirow{3}{*}{US}
% \multirow{6}{*}{Asia}

\begin{tabular}{llll}
\toprule
        Region & Codename &           Underlying index &                           ATH / Exchange \\
\midrule
        \multirow{9}{*}{Europe} &      AEX &          AEX (Netherlands) &  09:00 -- 17:30 CET / Euronext  \\
               &      CAC &             CAC40 (France) &               09:00 -- 17:30 CET / MONEP \\
               &      DAX &              DAX (Germany) &               09:00 -- 17:30 CET / Eurex \\
               &     FTSE &                  FTSE (UK) &               08:00 -- 17:30 GMT / LIFFE \\
               &      MIB &           FTSE MIB (Italy) &                 09:00 -- 17:30 CET / MIL \\
               &     IBEX &               IBEX (Spain) &                09:00 -- 17:30 CET / MEFF \\
               &    STOXX &        Euro STOXX (Europe) &               09:00 -- 17:30 CET / Eurex \\
               &     OMXS &  OMX Stockholm 30 (Sweden) &       09:00 -- 17:25 CET / OMX  \\
               &      SMI &          SMI (Switzerland) &               09:00 -- 17:25 CET / Eurex \\
\hline
            \multirow{3}{*}{US} &       ES &       S\&P 500, E-mini (US) &                 08:30 -- 15:15 CDT / CME \\
               &       DJ &     Dow Jones, E-mini (US) &                09:00 -- 15:15 CDT / CBOT \\
               &       NQ &        NASDAQ, E-mini (US) &                 08:30 -- 15:15 CDT / CME \\
\hline
          \multirow{6}{*}{Asia} &      HSI &      Hang Seng (Hong Kong) &                09:45 -- 12:30 HKT / HKFE \\
               &     HCEI &           HCEI (Hong Kong) &                09:45 -- 12:30 HKT / HKFE \\
               &   TAMSCI &         TAMSCI (Taiwan) &                08:45 -- 13:45 SGT / SEDT \\
               &    NIFTY &              NIFTY (India) &                10:00 -- 15:30 IST / NSEI \\
               &   NIKKEI &         Nikkei 225 (Japan) &                 12:30 -- 15:10 JST / OSA \\
               &    TOPIX &              TOPIX (Japan) &                 12:30 -- 15:10 JST / OSA \\
\hline
     Australia &      ASX &    S\&P/ASX 200 (Australia) &                09:50 -- 16:30 AEDT / SFE \\
\hline
 South America &  BOVESPA &           BOVESPA (Brazil) &               09:05 -- 17:15 BRT / SPCFE \\
\bottomrule
\end{tabular}

\end{center}
\end{table}

All Futures Contracts presented of Table~\ref{tb:contracts} are traded in different cycles with different expiration date. Moreover, for each Futures at every moment, 5--6 contracts with different maturities are traded simultaneously. Typically, for all Index Futures contracts, most of the trading activity is going at the so-called ``Front Month'' contract with the nearest maturity: in order to avoid settlement, financial investors ``roll over'' contracts to the next maturity typically one week before the expiration. 
At the rollover, the liquidity (measured in volume) of the contract that is going to expire is switched to the contract that will expire at the following quarter. 

\begin{figure}[t!]
  \centering
  \includegraphics[width=\textwidth]{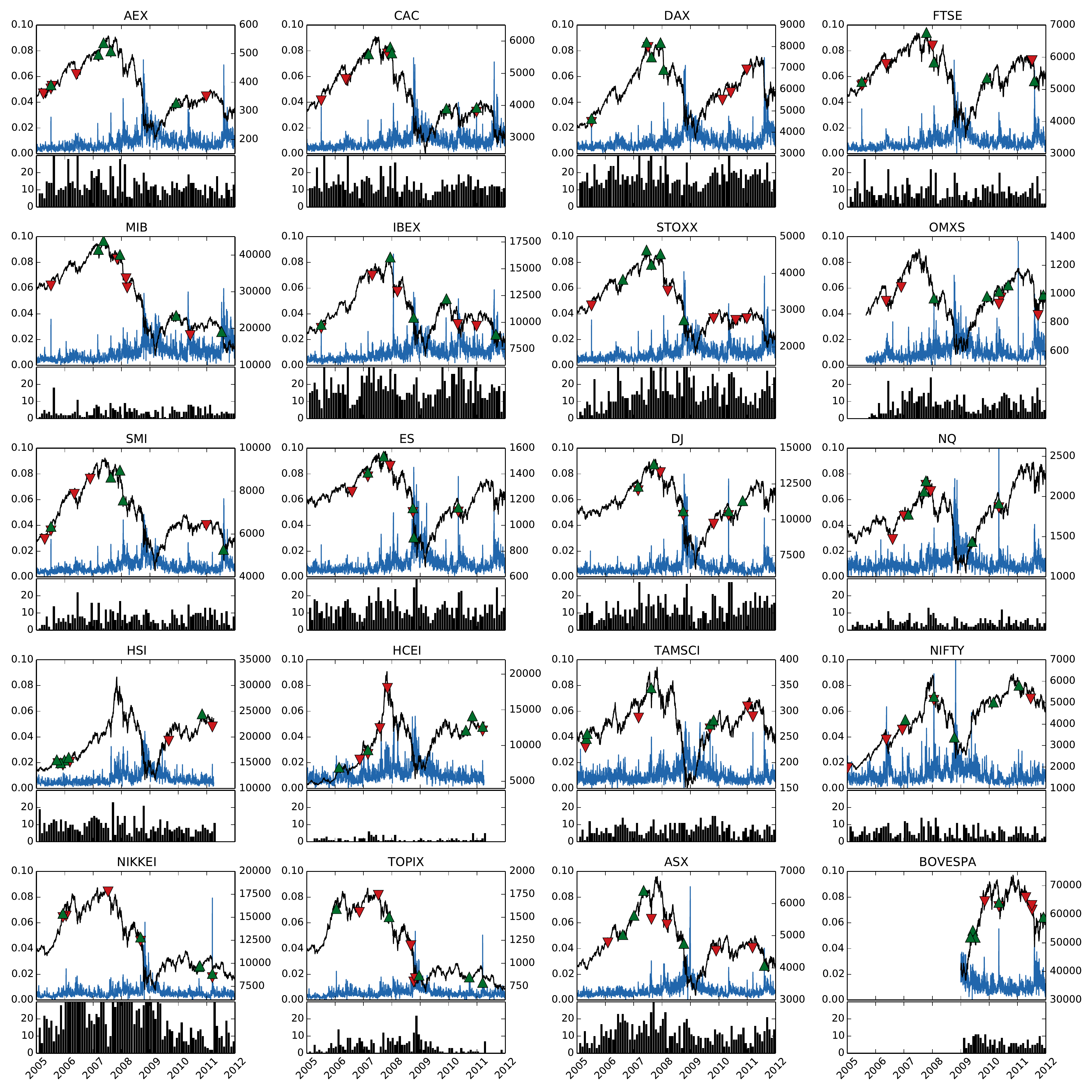}
  \caption{For each analyzed contract (Table~\ref{tb:contracts}), daily closing price (upper panel, black lines, right scale) and intraday volatility (upper panel, blue lines, left scale) are presented. 
Green up triangles and red down triangles denote respectively the 5 largest drawups and 
the 5 largest drawdowns for $\Delta t=30$ sec and $\epsilon_0=1$ (see Section~\ref{sec:dd}). 
The lower panel of each asset presents the number of ``largest'' drawdowns per month (drawdowns with normalized size larger than $\hat x_m$, see Section~\ref{sec:distrib_dd} and Table~\ref{tb:fit_dd_norm_returns}).}
\label{fig:price_vol}
\end{figure}

In order to construct a continuous time series from Futures Contracts with different maturities, we have ``rolled'' the previous contract to the next one on the second Thursdays of the expiration months. This approach is different from the traditional one ``rolling'' the Front Month  contract at expiration date. For any given date,
our approach amounts to selecting the contract with the largest daily volume. Note, however, that this ``rolling'' procedure, being well-suited for Futures contracts on financial indices, does not apply directly to other futures contracts such as commodity futures, where the physical delivery plays an important role (see for instance~\cite{Masteika2012} for a discussion of various methods of ``rolling'' procedures). Figure~\ref{fig:price_vol} presents the dynamics of daily closing prices of the analyzed Front Month contracts together with their intraday volatility estimated using \cite{GarmanKlass1980}'s method.

Before our analysis, we have cleaned the high-frequency data according to standard rules described in~\citep{Falkenberry2002,Brownlees2006,BarndorffNielsen2009}. Namely, we have (i) ignored all trades and quotes outside active trading hours (see Table~\ref{tb:descr}); (ii) ignored all recordings with transaction price, bid or ask equal to zero; (iii) deleted all quotes entries for which bid-ask spread is negative; (iv) deleted entries with the spread larger than 20 times the median spread of the day; (v) delete entries with corrected trades; (vi) deleted entries with prices above `ask' plus the bid-ask spread and entries with prices below `bid' minus bid-ask spread. Moreover, we ignored all trading days with gaps in data (due to issues on recording or low trading activity) longer than 5 minutes.

%===============================================================================
\section{Detection of drawdowns}\label{sec:dd}

Traditional measure of drawdowns (drawups) as sums of consecutive negative (positive) returns is very rigid and sensitive to noise: even tiny deviation of the price in the opposite direction will break large drawdowns into parts. Consider as an example a total peak-to-valley drop of $40\%$ of total duration of 10 days, with 
daily returns 
\begin{equation}
\{+0.8\%, -5\%. -3\%, -10\%, -2\%, +0.01\%, -8\%, -13\%, -3\%, -4\%, -2.01\%, +1.2\%\}~.
\label{rtjhetyngb}
\end{equation}
The second return $-5\%$ following the first positive return $+0.8\%$ determines the 
start of a drawdown. The last return $+1.2\%$ determines the end of the last drawdown.
In the standard definition, this series is characterised by two drawdowns
of $-20\%$, characterised by the followed return series,
\begin{eqnarray}
&&\{+0.8\%, -5\%. -3\%, -10\%, -2\%, +0.01\%\}, \nonumber \\
&&\{+0.01\%,  -8\%,  -13\%,  -3\%, -4\%, -2.01\%, +1.2\%\}~,
\label{rtjhewrthttyngb}
\end{eqnarray}
where we include the two positive returns at the boundaries to make clear the start and end
of the drawdowns. Does this makes sense? Following \cite{JohSorepsidd10},
we argue that investors gauging the market dynamics will be financially
and psychology hurt by the total loss of $-40\%$, and will be insensitive to the tiny
positive return $+0.01\%$ that technically defines two separate drawdowns.
As a risk measure, it is intuitive that the event to register is the total loss of $-40\%$.
This motivates the introduction of robust measures of drawdowns and we use
the so-called \emph{epsilon-drawdown} (\emph{$\epsilon$-drawdown}) measure introduced by \cite{JohSorepsidd10}. 
An $\epsilon$-drawdown is defined as a standard drawdown, except for the fact that
positive returns smaller than some defined threshold controlled by the parameter $\epsilon >0$
are considered as ``noise'' and do not end a drawdown run. Only when a positive
return occurs, which is larger than the threshold, is the drawdown deemed to end.
In the above example, taking a threshold of, say, $0.5\%$ leads to characterise
the series (\ref{rtjhetyngb}) as a single $\epsilon$-drawdown of amplitude $-40\%$,
which is a faithful embodiment of the realised losses of investors.
The two pure drawdowns of $20\%$, when analysed statistically for instance
via the distribution of their sizes that loses all information about their closeness,
paint a much milder picture of the true loss.
In practice, $\epsilon$ can be either a pre-defined constant or time-dependent. 
The second option is preferable to account for the clustering and memory effects 
of the volatility \citep{Cont2001}. For large-scale analyses and in order to 
compare different market regimes, the dynamics of volatility should be taken into account
for the choice of $\epsilon$. The most transparent way is to choose $\epsilon=\epsilon_0\sigma$, where $\sigma$ is a measure of the realized volatility estimated over a preceding time period
as discussed later and  $\epsilon_0$ is a constant. This time-adaptive choice for $\epsilon$ allows for a scaling 
of the tolerance in the definition of $\epsilon$-drawdown that takes into account the recent level of ``noise''.
For $\epsilon=0$, one recovers the classical definition of a drawdown (respectively, drawup) as a sequence of consecutive strictly negative (respectively, positive) returns.  

Technically, we define the sequence of drawups and drawdowns as follows. 
Consider the total time interval $[t_1, t_2]$. We first discretize it in  $N=[(t_2-t_1)/\Delta t]$ 
periods of length $\Delta t$, where the square brackets denote the floor function. 
This allows us to construct the discrete returns at time scale 
$\Delta t$ from the log-price series $p(t)=\log P(t)$ as
\begin{equation}\label{eq:ret}
  r_k=\log P(t_1+k\Delta t)-\log P(t_1+(k-1)\Delta t),\quad k=1,2, \dots, N~.
\end{equation}
The time $k=k_0=1$ is defined as a beginning of a drawup if $r_1>0$ and a drawdown if $r_1<0$. Then, for each $k>k_0$, we calculate the cumulative sum
\begin{equation}
  p_{k_0, k}=\sum_{i=k_0}^k r_i
\end{equation}
and test the largest deviation $\delta_{k_0,k}$ of the price trajectory from a previous extremum:
\begin{equation}\label{eq:delta}
  \delta_{k_0,k}=\begin{cases}
	  \displaystyle
	  p_{k_0, k}-\min_{k_0\leq i\leq k} p_{k_0,i} \qquad\text{for drawdowns},\\
	  \displaystyle
	  \max_{k_0\leq i\leq k} p_{k_0,i}-p_{k_0, k} \qquad\text{for drawups}.\\
  \end{cases}
\end{equation}
We stop the procedure when $\delta_{k_0,k}$ becomes larger than $\epsilon$:
\begin{equation}\label{eq:cond}
	\delta_{k_0,k}>\epsilon=\epsilon_0\sigma,
\end{equation}
where $\sigma$ is a measure of the volatility in the recent past and 
$\epsilon_0$ is a constant.

For a drawdown, the lowest price $k_1=\arg\min_{k_0\leq i\leq k} p_{k_0,i}$ is defined as its end and $k+1$ is defined as starting the following drawup. Respectively, for a drawup, the highest price $k_1=\arg\max_{k_0\leq i\leq k} p_{k_0,i}$ is defined as its end and $k+1$ as the beginning of the following drawdown. The procedure restarts from the time $k_1$, from which we compute the cumulative sum $p_{k_1, k}$ and the 
maximum deviation $\delta_{k_1,k}$, looking for the next value of $k$ that will satisfy  $\delta_{k_1,k}>\epsilon$. Then, the next extreme value of $\delta_{k_1,k}$ within $[k_1, k]$ provides us $k_2$, and so on. 

\begin{figure}[ht!]
  \centering
  \includegraphics[width=0.65\textwidth]{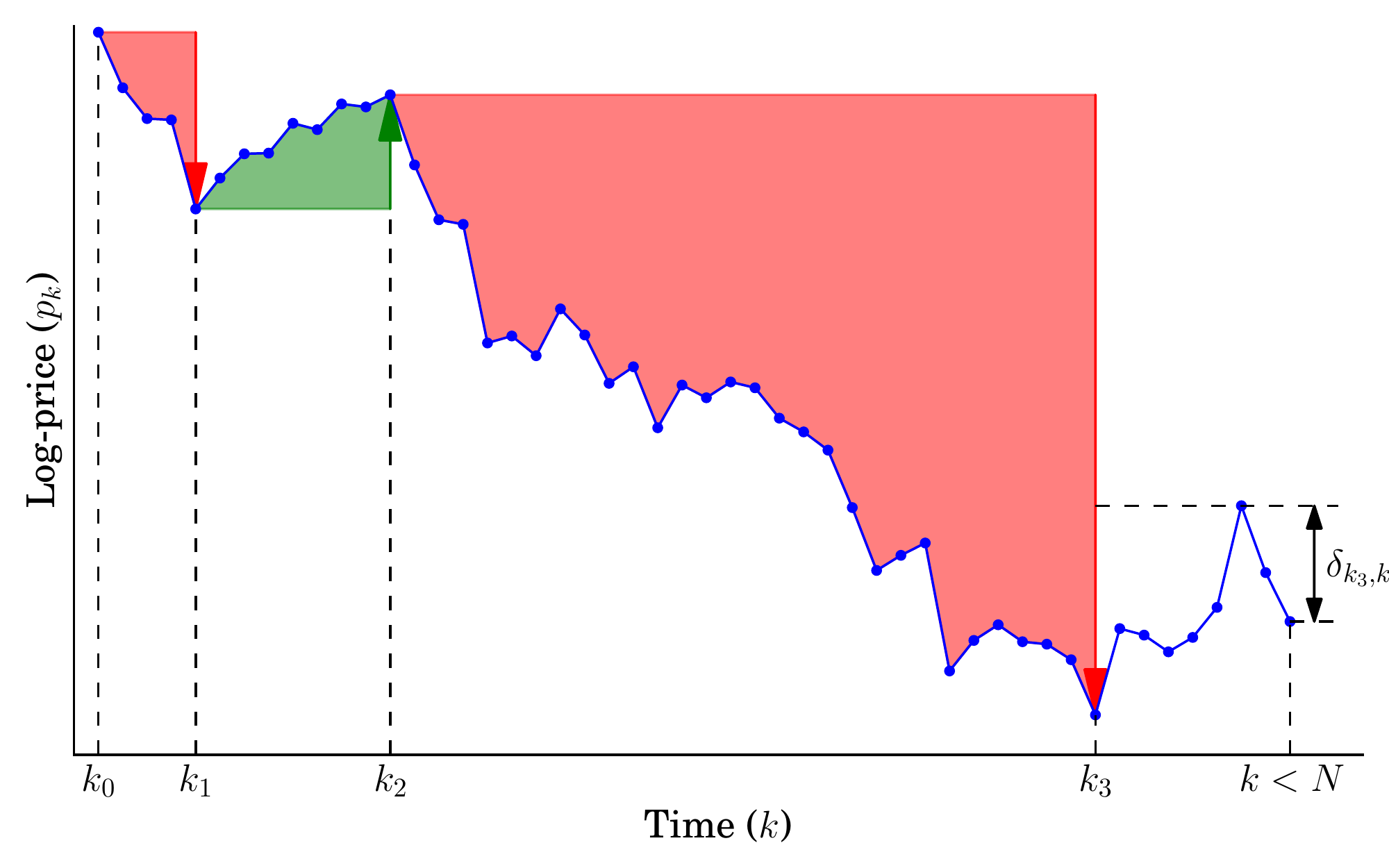}
  \caption{Illustration to the definition of $\epsilon$-drawups and $\epsilon$-drawdowns.}
\label{fig:drawdown_illustration}
\end{figure}

Figure~\ref{fig:drawdown_illustration} illustrates the procedure described above
and the fact that drawdowns are always followed by drawups and vice versa, by definition.
Drawdowns always start and end with negative returns, and correspondingly drawups always start and end with positive returns. It is important to notice that, similarly to other measures of trends, neither strict drawdowns nor $\epsilon$-drawdowns are causal, in the sense that at time $k$ it is impossible to say if the current drawdown (or drawup) is over or not. The classical definition ($\epsilon=0$) requires one-step look-ahead in order to conclude about the existence of the change of the trend and the end of the drawdown. The required look-ahead for $\epsilon$-drawdowns increases with $\epsilon$, and 
also exhibits some time variability through the realized volatility at the time of observation.

For the present analysis, we aggregate  tick-by-tick data within each day into 30-seconds bars ($\Delta t=30$ sec). The chosen size of $\Delta t$ results from a tradeoff. On the one hand,
$\Delta t$ should be long enough to reduce the microstructure noise (e.g. resulting from the bid-ask bounce) and to capture only systemic events while being insensitive to price drops and jumps due to ``fat-finger trades'' or the transient sparsity of the order book. On the another hand, $\Delta t$ should not
be too large so as to miss fast events that may happen on small time scales. 
For example, during one of the most dramatic intraday price fluctuations --- the so-called ``flash crash'' of May 6, 2010, that started on E-mini S\&P 500 futures contracts~\citep{FlashCrash2010_report}, and then almost instantly propagated to constituting stocks --- the price of E-mini Futures plunged by 5.7\% in 4 minutes and then recovered back by 4\% over the next 3 minutes. For instance, taking $\Delta t=15$~minutes would
prevent us from detecting any anomaly and would make use blind to this fast drawdown
followed by an equally fast drawup. As this flash crash is associated with
the largest intraday swing of the Dow Jones index (998.5 points or approximately 9\% in 5 minutes),
we must select parameters that ensure a suitable capture of the most salient developments of financial
markets.

In addition to $\Delta t$, the definition of $\epsilon$-drawdowns requires the specification
of $\epsilon$ defined by expression \eqref{eq:cond} in terms of the recent volatility $\sigma$.
Traditionally, the realized volatility at time $t$ in a running window of size $\tau$ is estimated as the standard deviation of the returns in the interval $[t-\tau, t)$. For an intraday estimation in
a time interval $[t_1, t_2]$, this estimation cannot be directly applied due to the
impact of the trading day opening, which is characterised by a very large transient volatility.
This leads to either truncate the window at its beginning in order to avoid or minimise the influence of the
opening period, which results in a smaller number of samples and thus higher variance of
the  initial estimates, or to include the overnight trading period prior to $t_1$, which 
amounts to mixing different regimes in one estimate. In order to be consistent within the trading day, 
for the $\sigma$ in expression \eqref{eq:cond}, 
we use the volatility calculated on returns of the previous trading day. 
Specifically, $\sigma$ is defined as the standard deviation of the log-returns $r_k$
defined by equation \eqref{eq:ret}:
\begin{equation}\label{eq:vol}
  \sigma^2=\frac 1N \sum_{k=0}^N r_k^2=
  \frac 1N \sum_{k=0}^N 
  \Big(\log P(t_1+k\Delta t)-\log P(t_1+(k-1)\Delta t)\Big)^2,
\end{equation}
where $N= [(t_2-t_1)/\Delta t]$. Note that expression~\eqref{eq:vol} is slightly different from the traditional definition of volatility $\sigma_{\Delta t}$, which is normalized by the time scale $\Delta t$: $\sigma_{\Delta t}=\sigma/\sqrt{\Delta t}$.

The intraday volatility is subjected to the so-called ``signature plot'' effect~\citep{AndersenBollerslev2000,Muzy2013hawkes,SaiSorepps14}: 
the volatility $\sigma_{\Delta t}$ is a decreasing function of
$\Delta t$. Moreover, in the presence of the bid-ask bounce,
decreasing $\Delta t$ tends to increase the negative autocorrelation of returns at 
the first lag. The choice of the parameter $\epsilon_0$ should 
thus be adapted to the time-scale $\Delta t$. For example, for $\Delta t$ of 
the order of few seconds, when most of the volatility $\sigma$ can be attributed to the bid-ask bounce, 
$\epsilon_0$ should be taken larger than for $\Delta t$ of the orders of minutes, in order to achieve the same level of aggregation. On the other hand, $\epsilon_0$ regulates the minimum (and also typical) size of the detected drawdowns, and plays a role of an effective ``scaling parameter''. 

For the analysis presented below, we have selected $\Delta t=30$ seconds and $\epsilon_0=1$. 
We have tested that our results are robust with respect to other values of these parameters
to within at least a factor $2$.

%===============================================================================
\section{Descriptive statistics of intraday drawdowns}\label{sec:descript_stat}

We characterize drawdowns (drawups) by the following properties:
\begin{itemize}
\item \emph{duration}: $\tau=(k_{end}-k_{start})\Delta t$, where $k_{start}$ and $k_{end}$ are time
index of the beginning and end of a drawdown (drawup). Since we work with
discrete times, durations are always multiple of $\Delta t$;
\item \emph{size}: $\Delta P=|P(t_1+k_{end}\Delta t)-P(t_1+k_{start}\Delta t)|$ (this is a multiple of a tick size);
\item \emph{return}: $r=|\log P(t_1+k_{end}\Delta t)-\log P(t_1+k_{start}\Delta t)|$;
\item \emph{normalized return}: $r^{norm}=r/\sigma$, where $\sigma$ is the volatility~\eqref{eq:vol} of the previous trading day;
\item \emph{speed}: $v=r/\tau$; 
\item \emph{normalized speed}: $v^{norm}=r^{norm}/\tau=v/\sigma$.
\end{itemize}
Over the 7 years of the period 2005--2011, all the studied markets have passed through different regimes (see Figure~\ref{fig:price_vol}): low-volatile period of 2005--2007, increase of uncertainty in 2008 and volatility burst during the peak of the sub-prime crisis in October 2008. After the relaxation to the pre-crisis level by the second half of 2009--early 2010, the volatility spiked again in the mid-2010 and second half of 2010 due to the Eurozone debt crisis. For example, for E-mini S\&P 500 Futures Contracts, the highest intraday volatility over the whole period ($0.083$ on October 10, 2008) is almost 50 times larger than the lowest ($0.0015$ on February 15, 2007). Thus, normalizing drawdowns by the instantaneous volatility is essential
to obtain meaningful results aggregated 
over such a long period and especially across multiple assets.

\begin{table}[t!]
\vspace{-2ex}
\caption{Descriptive statistics of drawups and drawdowns for $\Delta t=30$ sec and $\epsilon_0=1$. The table summarizes the number of detected drawdowns and drawups (``Count''), as well as the median, 90\% quantile ($Q_{90\%}$) and maximal value of duration ($\tau$), size ($\Delta P$), normalized return ($r^{norm}$) and normalized speed ($v^{norm}$) of drawdowns and drawups.  
}

\label{tb:descr}
\begin{center}
%\footnotesize
\scriptsize
\renewcommand{\arraystretch}{0.8}

(a) Drawdowns
\makebox[\textwidth][c]{
\begin{tabular}{l|r|rrr|rrr|rrr|rrr}
\toprule
 \multirow{2}{*}{Codename} &  \multirow{2}{*}{Count} & 
 \multicolumn{3}{c|}{$\tau$, sec} & \multicolumn{3}{c|}{$|\Delta P|$} & 
 \multicolumn{3}{c|}{$|r^{norm}|$} & \multicolumn{3}{c}{$|v^{norm}|$} \\
 %\cline{3-14}
&& Median & $Q_{90\%}$ & Max & Median & $Q_{90\%}$ & Max &
Median & $Q_{90\%}$ & Max & Median & $Q_{90\%}$ & Max \\
\midrule
     AEX &  156981 &      120 &    360 &     4860 &    0.25 &    0.75 &     8.75 &   2.49 &  6.48 &   68.72 &  0.024 &  0.060 &  1.079 \\
     CAC &  159096 &      120 &    360 &     6450 &    3.00 &    9.00 &   123.00 &   2.53 &  6.37 &  215.00 &  0.024 &  0.059 &  0.896 \\
     DAX &  161094 &      120 &    360 &     3030 &    4.50 &   13.00 &   141.00 &   2.49 &  6.19 &  119.51 &  0.023 &  0.058 &  0.932 \\
    FTSE &  181065 &      120 &    360 &     3150 &    3.50 &   10.50 &   138.00 &   2.51 &  6.23 &  176.47 &  0.024 &  0.057 &  1.164 \\
     MIB &  151537 &      120 &    360 &     4800 &   20.00 &   55.00 &   700.00 &   2.45 &  6.15 &   88.45 &  0.024 &  0.059 &  0.666 \\
    IBEX &  157868 &      120 &    360 &     3720 &    9.00 &   25.00 &   580.00 &   2.47 &  6.11 &  147.75 &  0.024 &  0.059 &  0.876 \\
   STOXX &  194997 &       90 &    300 &     4170 &    2.00 &    6.00 &    80.00 &   2.10 &  5.12 &   78.21 &  0.026 &  0.056 &  0.769 \\
    OMXS &  115381 &      120 &    390 &     5100 &    0.75 &    2.25 &    20.75 &   2.44 &  6.17 &   52.25 &  0.024 &  0.061 &  0.610 \\
     SMI &  155845 &      120 &    360 &     6480 &    4.00 &   12.00 &   143.00 &   2.46 &  6.32 &  121.38 &  0.024 &  0.059 &  1.349 \\
      ES &  135324 &      120 &    330 &     5670 &    1.00 &    2.50 &    59.50 &   2.40 &  5.63 &  132.62 &  0.024 &  0.057 &  0.553 \\
      DJ &  121924 &      120 &    330 &     2370 &    7.00 &   20.00 &   551.00 &   2.47 &  5.97 &  165.05 &  0.024 &  0.058 &  0.688 \\
      NQ &  134759 &      120 &    330 &     2700 &    1.50 &    4.50 &   120.00 &   2.42 &  5.97 &  149.19 &  0.024 &  0.057 &  0.498 \\
     HSI &   47294 &      120 &    330 &     3000 &   18.00 &   56.00 &   581.00 &   2.55 &  6.08 &   62.51 &  0.024 &  0.055 &  0.468 \\
    HCEI &   46411 &      120 &    330 &     2250 &   14.00 &   45.00 &   629.00 &   2.50 &  6.27 &   44.34 &  0.024 &  0.056 &  0.343 \\
  TAMSCI &  100446 &       90 &    330 &     4770 &    0.30 &    0.70 &     6.70 &   2.27 &  5.69 &   41.45 &  0.024 &  0.055 &  0.740 \\
   NIFTY &   99165 &      120 &    330 &     2250 &    4.31 &   12.50 &   199.00 &   2.63 &  6.62 &   67.75 &  0.025 &  0.056 &  0.665 \\
  NIKKEI &   72177 &       60 &    240 &     1770 &   10.00 &   20.00 &   290.00 &   1.70 &  3.68 &   27.77 &  0.029 &  0.059 &  0.457 \\
   TOPIX &   65737 &       90 &    270 &     2010 &    0.50 &    2.00 &    35.50 &   1.66 &  4.67 &   30.90 &  0.027 &  0.053 &  0.263 \\
     ASX &  119717 &      120 &    390 &     2700 &    3.00 &    9.00 &    75.00 &   2.51 &  6.27 &   36.88 &  0.024 &  0.057 &  0.420 \\
 BOVESPA &   58885 &      120 &    360 &     3180 &   60.00 &  160.00 &  1600.00 &   2.52 &  6.21 &   50.42 &  0.023 &  0.057 &  0.746 \\
\bottomrule
\end{tabular}
}
(b) Drawups
\makebox[\textwidth][c]{
\begin{tabular}{l|r|rrr|rrr|rrr|rrr}
\toprule
 \multirow{2}{*}{Codename} &  \multirow{2}{*}{Count} & 
 \multicolumn{3}{c|}{$\tau$, sec} & \multicolumn{3}{c|}{$\Delta P$} & 
 \multicolumn{3}{c|}{$r^{norm}$} & \multicolumn{3}{c}{$v^{norm}$} \\
 %\cline{3-14}
&& Median & $Q_{90\%}$ & Max & Median & $Q_{90\%}$ & Max &
Median & $Q_{90\%}$ & Max & Median & $Q_{90\%}$ & Max \\
\midrule
    AEX &  157018 &      120 &    390 &     4200 &    0.25 &    0.750 &    15.25 &   2.50 &  6.48 &   48.72 &  0.024 &  0.060 &  0.830 \\
     CAC &  159127 &      120 &    360 &     3750 &    3.00 &    9.000 &   179.50 &   2.54 &  6.37 &  157.99 &  0.024 &  0.058 &  0.699 \\
     DAX &  161102 &      120 &    360 &     3390 &    4.50 &   13.000 &   265.50 &   2.51 &  6.24 &   52.53 &  0.023 &  0.057 &  1.004 \\
    FTSE &  181136 &      120 &    360 &     4590 &    3.50 &   10.500 &   219.50 &   2.51 &  6.23 &   79.00 &  0.024 &  0.056 &  2.634 \\
     MIB &  151527 &      120 &    390 &     4680 &   20.00 &   55.000 &   810.00 &   2.45 &  6.11 &   50.36 &  0.023 &  0.058 &  0.789 \\
    IBEX &  157985 &      120 &    360 &     3660 &    9.00 &   25.000 &   646.00 &   2.47 &  6.12 &   52.00 &  0.024 &  0.058 &  0.657 \\
   STOXX &  195052 &       90 &    300 &     4980 &    2.00 &    6.000 &   131.00 &   2.11 &  5.13 &   59.66 &  0.026 &  0.055 &  0.698 \\
    OMXS &  115430 &      120 &    390 &     4620 &    0.75 &    2.250 &    39.00 &   2.45 &  6.12 &   43.86 &  0.024 &  0.061 &  0.631 \\
     SMI &  155897 &      120 &    390 &     4050 &    4.00 &   12.000 &   293.00 &   2.47 &  6.31 &   59.09 &  0.024 &  0.058 &  0.842 \\
      ES &  135359 &      120 &    330 &     3060 &    1.00 &    2.500 &    41.25 &   2.41 &  5.64 &   83.32 &  0.024 &  0.057 &  1.187 \\
      DJ &  121968 &      120 &    330 &     2880 &    7.00 &   20.000 &   264.00 &   2.49 &  5.99 &   72.97 &  0.024 &  0.057 &  1.518 \\
      NQ &  134784 &      120 &    330 &     3270 &    1.50 &    4.500 &    76.50 &   2.44 &  5.97 &   96.25 &  0.024 &  0.057 &  0.803 \\
     HSI &   47410 &      120 &    330 &     2760 &   18.00 &   56.000 &   409.00 &   2.56 &  6.09 &   46.75 &  0.024 &  0.054 &  0.456 \\
    HCEI &   46497 &      120 &    360 &     2220 &   14.00 &   44.000 &   402.00 &   2.51 &  6.23 &   44.19 &  0.024 &  0.056 &  0.366 \\
  TAMSCI &  100481 &       90 &    330 &     5640 &    0.30 &    0.700 &    14.80 &   2.27 &  5.65 &  127.45 &  0.024 &  0.055 &  0.555 \\
   NIFTY &   99148 &      120 &    330 &     1890 &    4.40 &   12.365 &   204.65 &   2.70 &  6.56 &   48.82 &  0.025 &  0.055 &  0.680 \\
  NIKKEI &   72199 &       60 &    240 &     1890 &   10.00 &   20.000 &   270.00 &   1.70 &  3.64 &   31.15 &  0.029 &  0.059 &  0.557 \\
   TOPIX &   65815 &       90 &    270 &     2010 &    0.50 &    2.000 &    23.00 &   1.68 &  4.63 &   35.19 &  0.027 &  0.053 &  0.321 \\
     ASX &  119818 &      120 &    390 &     5880 &    3.00 &    9.000 &   109.00 &   2.52 &  6.20 &   44.05 &  0.024 &  0.056 &  0.351 \\
 BOVESPA &   58900 &      120 &    360 &     2730 &   60.00 &  160.000 &  1600.00 &   2.54 &  6.17 &   51.96 &  0.023 &  0.057 &  0.626 \\
\bottomrule
\end{tabular}
}
\end{center}
\end{table}

Table~\ref{tb:descr} presents descriptive statistics of durations, sizes, normalized returns and speeds of drawups and drawdowns detected for $\Delta t=30$ sec and $\epsilon_0=1$ on the entire interval of 2005--2011 for different contracts (for BOVESPA, we considered the period 2009--2011). Changes in $\Delta t$ and $\epsilon_0$ modify the descriptive statistics as follows. Increasing  $\Delta t$ or $\epsilon_0$ decreases proportionally the number of drawdowns and drawups, and magnifies their durations
$\tau$ together with their sizes $\Delta P$ and their normalized returns $r^{norm}$. 
For instance, for $\Delta t=2$ min and $\epsilon_0=2$, the typical duration $\tau$  
is roughly 14--15 times larger than for $\Delta t=30$ sec and $\epsilon_0=0.5$.
While changes of $\Delta P$ are proportional to changes in $\Delta t$ and $\epsilon_0$, 
the normalized returns $r^{norm}$ only gradually increase with $\Delta t$ and, at the same time,
are much more sensitive to $\epsilon_0$. 
Interestingly, the normalized speed $v^{norm}$ is almost insensitive to changes in $\epsilon_0$, but decreases with increase of $\Delta t$ (i.e. larger drawdowns are typically slower than shorter drawdowns). This decrease is almost proportional for most of the distribution of normalized speeds, except
in its tail where a different scaling holds.
Such non-trivial scaling can be interpreted as due to the multifractal properties of the price dynamics~\citep{ArneodoMuzySornette1998,Calvet2002review,Muzy2005} resulting
from the interplay between long-term memory in the system together with nonlinear amplification~\citep{FilimonovSornette2011_SEMF}.

Table~\ref{tb:descr} shows that the statistics for drawdowns and drawups are almost identical for each asset and for all analyzed parameters, except the statistics of extreme (maximal) values.
Typically (for developed European and US markets), extreme drawdowns are larger than extreme drawups in terms of normalized returns $r^{norm}$, which concurs with the empirical evidence of gain-loss asymmetry~\citep{Cont2001,Jensen2003}. However, for TAMSCI (Taiwan), the extreme drawup of 5\% at the opening of the market on September 10, 2009 ($r^{norm}=127.4$) is more than twice larger than the maximum observed drawdown ($r^{norm}=57.7$). For NIKKEI, TOPIX (Japan) and BOVESPA (Brazil), the largest drawup is only slightly larger than the most extreme drawdown.

We observe that both median and 90\%-quantiles (as well as mean not reported in the table) of the durations $\tau$ of drawdowns and drawups are almost equal to each other for all Futures contracts, except NIKKEI and TOPIX. Though the price of contracts vary in a wide range and so does the sizes $\Delta P$ of drawdowns, the statistics (mean, median and quantiles) of normalized returns $r^{norm}$ are very similar across contracts. At the same time, maximal values of the normalized returns differ by a large factor: the smallest extreme value ($r^{norm}=44.3$ for HCEI) is 4.8 times smaller than the largest one ($r^{norm}=215.0$ for CAC). Finally, mean and quantile statistics for normalized speeds $v^{norm}$ agree almost perfectly across all contracts.

We also note the exceptional properties of Japanese markets (contracts NIKKEI and TOPIX), which exhibit the shortest (minimal values of $\tau$) and the smallest (minimal values of $r^{norm}$) drawdowns and drawups among all analyzed contracts. Moreover, while the normalized speed $v^{norm}$ of drawdowns and drawups are typically close to each other across different contracts, the normalized speed of the fastest drawdown on TOPIX Futures Contracts is at least twice smaller than that of any other analyzed contract.

%===============================================================================
\section{Distribution of normalized size of drawdowns}\label{sec:distrib_dd}

\begin{figure}[t!]
  \centering
  \includegraphics[width=\textwidth]{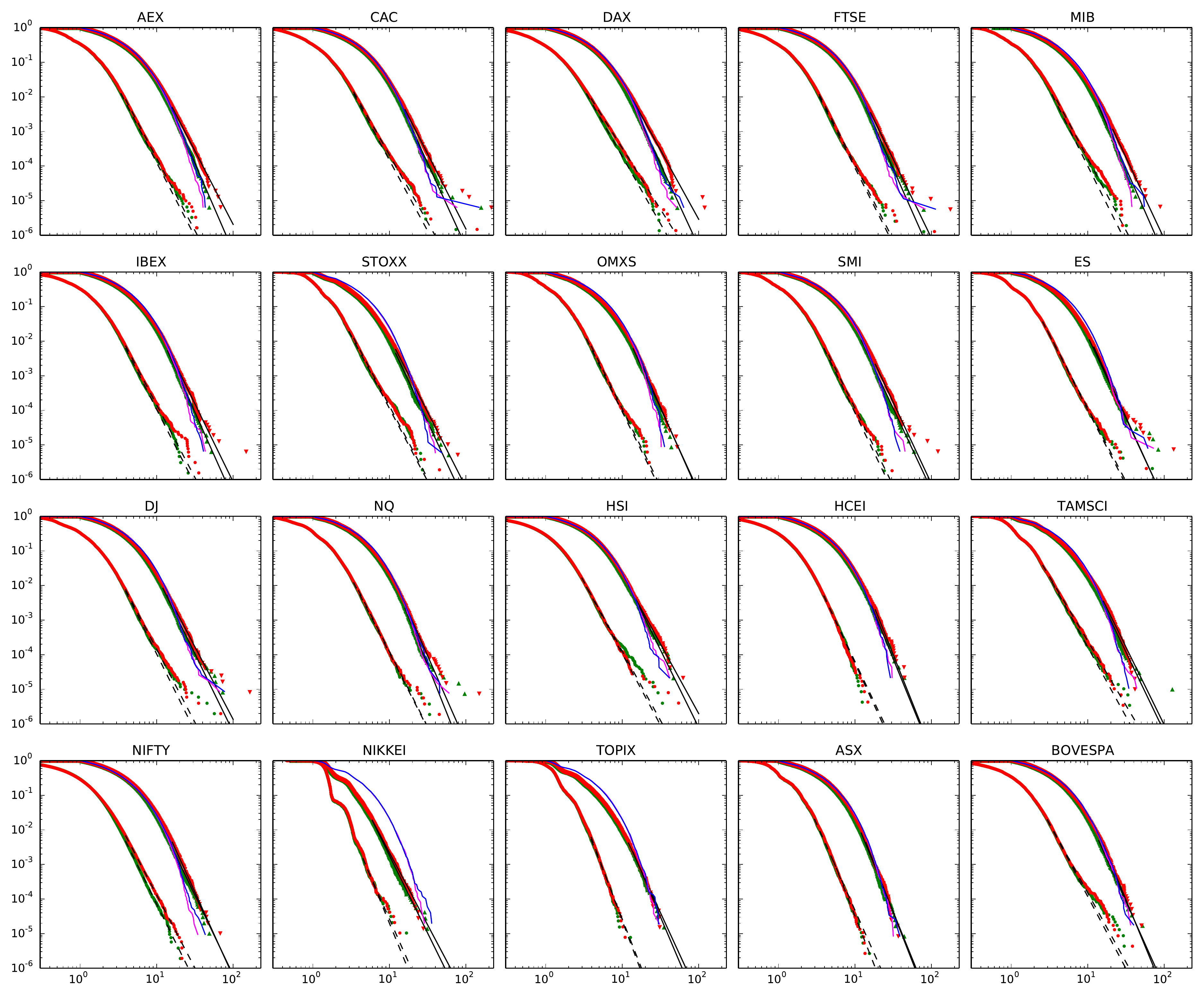}
  \caption{Complementary cumulative distribution function (ccdf) for the normalized returns of drawdowns (red down triangles) and drawups (green up triangles) for $\Delta t=30$ sec and $\epsilon_0=1$. Blue and magenta lines depict the distributions of drawdowns and drawups of the null model constructed by reshuffled returns. Black straight lines correspond to power law fits of the tails of the distributions of drawdowns (see Table~\ref{tb:fit_dd_norm_returns}). Red and green dots lines show the distributions of normalized returns at $\Delta t=30$ sec. Dashed black lines correspond to power law fits of the tails of these distributions of returns (see Table~\ref{tb:fit_dd_norm_returns}). }
\label{fig:ccdf}
\end{figure}

Figure~\ref{fig:ccdf} shows the complementary cumulative distribution functions (ccdf) of the normalized returns $r^{norm}$ of the drawdowns and drawups defined at the top of the previous 
section \ref{sec:descript_stat}, for $\Delta t=30$ sec and $\epsilon_0=1$, together with the distributions of negative and positive price log-returns for the value $\Delta t=30$. Together with the empirical distributions, we  present the distributions of returns of drawdowns for the null model constructed by reshuffling log-returns within the active trading hours of each trading day independently. This reshuffling destroys all temporal correlations while keeping unchanged the marginal distributions of log-returns as well as the secular dynamics of the intraday volatility (see Figure~\ref{fig:price_vol}).

One can observe in Figure~\ref{fig:ccdf} that the tails of the distributions of drawdowns and drawups deviate from those obtained for the null model. While the drawdowns in the null model follow approximately a Weibull distribution with  shape parameter estimated around $0.9$, the real data is much better characterised by a fatter power law tail.
This is evidence that time dependence between returns play an important role in the directional price movements and especially in the occurrence of large drawdowns and drawups. Interestingly, for some of the analyzed contracts (CAC, FTSE, ES and DJ), one can observe significant deviations 
in the null model distribution that are associated with a few extreme events.
They correspond to drawdowns whose size is essentially controlled by a single 
return of enormous magnitude: as seen from Figure~\ref{fig:ccdf}, the sizes of these drawdowns are identical to the sizes of the return outliers.

In order to quantify the power law approximation of the tails,
\begin{equation}\label{eq:pl_ccdf}
    F(x)=\mathrm{Pr}\left[r^{norm} >x\right] = \left(\frac{x_{m}}{x}\right)^{\alpha},\quad x\ge x_m
\end{equation}
we have employed the framework proposed in~\citep{Clauset2009}. Assuming a known value of $x_m$,  the (Hill) maximum likelihood estimator (MLE) yields the well-known closed form expression for the exponent $\hat \alpha$:
\begin{equation}\label{eq:mle}
   \hat \alpha=N\cdot\left[\sum_{i=1}^N\ln\frac{x_i}{x_m}\right]^{-1},
\end{equation}
where $N$ is the number of observations in the tail (i.e. such that $x_i\ge x_m$). The standard error on $\hat\alpha$ is derived from the width of the likelihood maximum~\citep{SorknoRO96,Newman2005,Clauset2009}:
\begin{equation}\label{eq:mle_error}
    \sigma_{\hat\alpha} \simeq \frac{\hat\alpha}{\sqrt{N}}
\end{equation}
Following~\citet{Clauset2009}, we have scanned different values of $x_m$, fitting the distribution~\eqref{eq:pl_ccdf} and calculating the Kolmogorov-Smirnov distance~\citep{deGrootSchervish2011_Prob_n_Stats} between the cumulative distribution function 
(cdf) of the data and fitted model:
\begin{equation}\label{eq:KS_dist}
    D = \max_{x\ge x_m}|S(x)-F(x)|,
\end{equation}
where $S(x)$ is the empirical cdf for the observed data and $F(x)$ is given by~\eqref{eq:pl_ccdf}. As an alternative test, we have used the Anderson-Darling distance~\citep{deGrootSchervish2011_Prob_n_Stats}, which is more sensitive to the deviations in the tails of the distributions
\begin{equation}\label{eq:A2_dist}
    A^2 = N\int_{x_m}^\infty \frac{(S(x)-F(x))^2}{F(x)(1-F(x))}dF(x)
    %A^2 
    = -N -\sum_{i=1}^N\frac{2i-1}{N}\Big[
    \ln F(x_i)+\ln (1-F(x_{N+1-i}))
    \Big],
\end{equation}
where the input data is assumed to be ordered ($x_m\leq x_1<x_2<\dots<x_N$).
We then used the value of $\hat x_m$ that minimizes $D$ (or $A^2$) as an estimate of the lower bound $x_m$. An illustration of the power law fits of the tails of the distributions of drawups and drawdowns is presented in Figure~\ref{fig:ccdf}.

Table~\ref{tb:fit_dd_norm_returns} summarises the results obtained by fitting the power law distribution~\eqref{eq:pl_ccdf} to the empirical ccdf of the normalized returns ($r^{norm}$) of drawdowns and drawups using the Kolmogorov-Smirnov distance~\eqref{eq:KS_dist} for the lower bound detection. The Anderson-Darling distance~\eqref{eq:A2_dist} provides values similar to those reported in Table~\ref{tb:fit_dd_norm_returns} except for the contracts ES, NQ, HCEI, NIKKEI, TOPIX and BOVESPA 
for which the more conservative $A^2$ distance cuts off too much of the data by estimating 
a lower bound $\hat x_m\gtrsim 20$. The resulting dataset contains less than 100 points, which prevents a robust estimation. This behavior resulting from the use of the $A^2$ distance was previously discussed in~\citet{Clauset2009}. 
Let us mention that alternative methods have recently been introduced that improve on or complement  
the maximum likelihood (Hill) estimation of exponent $\alpha$
together with the selection of the lower bound $x_m$ using the Kolmogorov-Smirnov or Anderson-Darling distances.
These methods include \citep{Carpenter2013,Deluca2013,Wager2014}), each of which 
present advantages over the Hill estimator but have also their own limitations.

\begin{table}[hbt!]
%\vspace{-2ex}
\caption{Estimates of the lower boundary $\hat x_m$ and of the exponent $\hat \alpha$ of the power law fits~\eqref{eq:pl_ccdf} of the distributions of normalized returns $r^{norm}$ of the drawdown and drawup for $\Delta t=30$ sec and $\epsilon_0=1$ and of the normalized positive and negative price log-returns for $\Delta t=30$ sec. The exponents $\hat \alpha$ are presented together with the estimation of its standard error~\eqref{eq:mle_error}; the number of observations in the power law tail ($N_{x\ge \hat x_m}$) is also reported.
}
\label{tb:fit_dd_norm_returns}

\begin{center}
%\scriptsize
\footnotesize
\renewcommand{\arraystretch}{0.8}

(a) Drawdowns and negative returns
%\begin{tabular}{l | rrrrr | rrrrr}
\begin{tabular}{l | rrr | rrr}
\toprule
 \multirow{2}{*}{Codename} &  \multicolumn{3}{c|}{Drawdowns} & \multicolumn{3}{c}{Negative returns} \\
 & \multicolumn{1}{c}{$\hat x_m$} & \multicolumn{1}{c}{$\hat\alpha$} & $N_{x\ge \hat x_m}$ 
 & \multicolumn{1}{c}{$\hat x_m$} & \multicolumn{1}{c}{$\hat\alpha$} & $N_{x\ge \hat x_m}$ \\
\midrule
AEX     &   14.83 &  4.25 $\pm$ 0.13 &    1068 &     3.72 &  4.18 $\pm$ 0.05 &     6342 \\
CAC     &   14.38 &  4.34 $\pm$ 0.13 &    1076 &     3.06 &  3.90 $\pm$ 0.03 &    14311 \\
DAX     &   13.32 &  4.00 $\pm$ 0.10 &    1481 &     3.45 &  3.53 $\pm$ 0.03 &    10226 \\
FTSE    &   15.38 &  4.56 $\pm$ 0.17 &     752 &     3.23 &  4.33 $\pm$ 0.04 &    11505 \\
MIB     &   18.43 &  4.71 $\pm$ 0.26 &     337 &     3.49 &  4.16 $\pm$ 0.05 &     6844 \\
IBEX    &   12.70 &  4.51 $\pm$ 0.12 &    1493 &     3.68 &  4.20 $\pm$ 0.05 &     6138 \\
STOXX   &   11.60 &  4.34 $\pm$ 0.12 &    1326 &     3.09 &  4.19 $\pm$ 0.04 &     9590 \\
OMXS    &   14.71 &  5.00 $\pm$ 0.20 &     655 &     3.78 &  4.47 $\pm$ 0.09 &     2724 \\
SMI     &   16.25 &  4.71 $\pm$ 0.19 &     605 &     4.12 &  4.48 $\pm$ 0.08 &     3386 \\
ES      &   11.73 &  4.85 $\pm$ 0.15 &    1041 &     2.56 &  4.24 $\pm$ 0.03 &    17897 \\
DJ      &   12.40 &  4.19 $\pm$ 0.13 &    1042 &     3.86 &  4.23 $\pm$ 0.07 &     4127 \\
NQ      &   16.18 &  5.01 $\pm$ 0.27 &     358 &     3.76 &  4.36 $\pm$ 0.06 &     4633 \\
HSI     &   11.07 &  4.03 $\pm$ 0.16 &     662 &     2.89 &  4.00 $\pm$ 0.06 &     4561 \\
HCEI    &   18.18 &  5.54 $\pm$ 0.56 &      99 &     3.40 &  4.80 $\pm$ 0.10 &     2332 \\
TAMSCI  &   14.27 &  4.50 $\pm$ 0.19 &     591 &     2.37 &  3.72 $\pm$ 0.03 &    15115 \\
NIFTY   &   15.85 &  4.88 $\pm$ 0.23 &     456 &     3.84 &  4.21 $\pm$ 0.07 &     3850 \\
NIKKEI  &    5.56 &  4.18 $\pm$ 0.10 &    1841 &     5.14 &  5.00 $\pm$ 0.59 &       71 \\
TOPIX   &   11.67 &  4.81 $\pm$ 0.27 &     321 &     3.69 &  5.93 $\pm$ 0.18 &     1125 \\
ASX     &   12.42 &  5.67 $\pm$ 0.18 &     972 &     5.17 &  5.62 $\pm$ 0.25 &      490 \\
BOVESPA &   15.92 &  5.37 $\pm$ 0.36 &     221 &     2.69 &  4.01 $\pm$ 0.05 &     6758 \\
\bottomrule
\end{tabular}

%\vspace{0.3cm}
(b) Drawups and positive returns
%\vspace{0.1cm}

%\begin{tabular}{l | rrrrr | rrrrr}
\begin{tabular}{l | rrr | rrr}
\toprule
 \multirow{2}{*}{Codename} &  \multicolumn{3}{c|}{Drawups} & \multicolumn{3}{c}{Positive returns} \\
 & \multicolumn{1}{c}{$\hat x_m$} & \multicolumn{1}{c}{$\hat\alpha$} & $N_{x\ge \hat x_m}$ 
 & \multicolumn{1}{c}{$\hat x_m$} & \multicolumn{1}{c}{$\hat\alpha$} & $N_{x\ge \hat x_m}$ \\
\midrule
AEX     &   16.74 &  5.10 $\pm$ 0.23 &     486 &     3.37 &  4.27 $\pm$ 0.05 &     8065 \\
CAC     &   14.05 &  4.82 $\pm$ 0.16 &     950 &     3.17 &  4.09 $\pm$ 0.04 &    11218 \\
DAX     &   13.50 &  4.77 $\pm$ 0.15 &    1059 &     3.37 &  3.89 $\pm$ 0.04 &     9585 \\
FTSE    &   12.58 &  4.99 $\pm$ 0.13 &    1481 &     3.05 &  4.40 $\pm$ 0.04 &    12992 \\
MIB     &   13.70 &  5.04 $\pm$ 0.17 &     876 &     3.18 &  4.42 $\pm$ 0.05 &     8339 \\
IBEX    &   12.56 &  4.96 $\pm$ 0.13 &    1363 &     3.47 &  4.34 $\pm$ 0.05 &     6814 \\
STOXX   &   10.61 &  4.74 $\pm$ 0.12 &    1584 &     2.56 &  4.21 $\pm$ 0.03 &    18461 \\
OMXS    &   12.86 &  4.82 $\pm$ 0.15 &    1013 &     3.00 &  4.51 $\pm$ 0.06 &     6721 \\
SMI     &   11.54 &  4.72 $\pm$ 0.10 &    2073 &     3.42 &  4.60 $\pm$ 0.06 &     6587 \\
ES      &   10.41 &  4.75 $\pm$ 0.12 &    1525 &     2.54 &  4.33 $\pm$ 0.03 &    18327 \\
DJ      &   11.16 &  4.52 $\pm$ 0.12 &    1431 &     3.73 &  4.46 $\pm$ 0.07 &     4368 \\
NQ      &   14.63 &  5.55 $\pm$ 0.25 &     487 &     3.07 &  4.35 $\pm$ 0.04 &    10172 \\
HSI     &   13.91 &  4.48 $\pm$ 0.28 &     259 &     3.08 &  4.13 $\pm$ 0.07 &     3456 \\
HCEI    &   15.14 &  5.43 $\pm$ 0.39 &     191 &     3.39 &  4.65 $\pm$ 0.10 &     2279 \\
TAMSCI  &   14.31 &  4.61 $\pm$ 0.22 &     457 &     2.95 &  4.02 $\pm$ 0.05 &     6551 \\
NIFTY   &   11.41 &  4.61 $\pm$ 0.12 &    1406 &     3.55 &  4.47 $\pm$ 0.07 &     3891 \\
NIKKEI  &    5.88 &  4.49 $\pm$ 0.12 &    1425 &     3.91 &  5.42 $\pm$ 0.31 &      313 \\
TOPIX   &   13.36 &  5.22 $\pm$ 0.39 &     180 &     3.63 &  5.65 $\pm$ 0.17 &     1051 \\
ASX     &   11.94 &  5.45 $\pm$ 0.17 &    1029 &     2.82 &  4.98 $\pm$ 0.05 &     9786 \\
BOVESPA &   12.62 &  4.93 $\pm$ 0.23 &     460 &     2.78 &  4.17 $\pm$ 0.06 &     5669 \\
\bottomrule
\end{tabular}

\end{center}
\end{table}

Table~\ref{tb:fit_dd_norm_returns} summarizes the values of the lower boundary $\hat x_m$ and exponent $\hat \alpha$ of the power law fits of the distributions of the normalized returns ($r^{norm}$) for drawdowns and drawups for $\Delta t=30$ sec and $\epsilon_0=1$. Our tests show that, when increasing both $\Delta t$ and $\epsilon_0$, the exponent $\hat\alpha$ slightly increases. However, overall, the exponent estimates are consistent for all values of $\Delta t$ and $\epsilon_0$ and coincide with a relatively good precision between drawups and drawdowns. 

\begin{figure}[htb!]
  \centering
  \includegraphics[width=\textwidth]{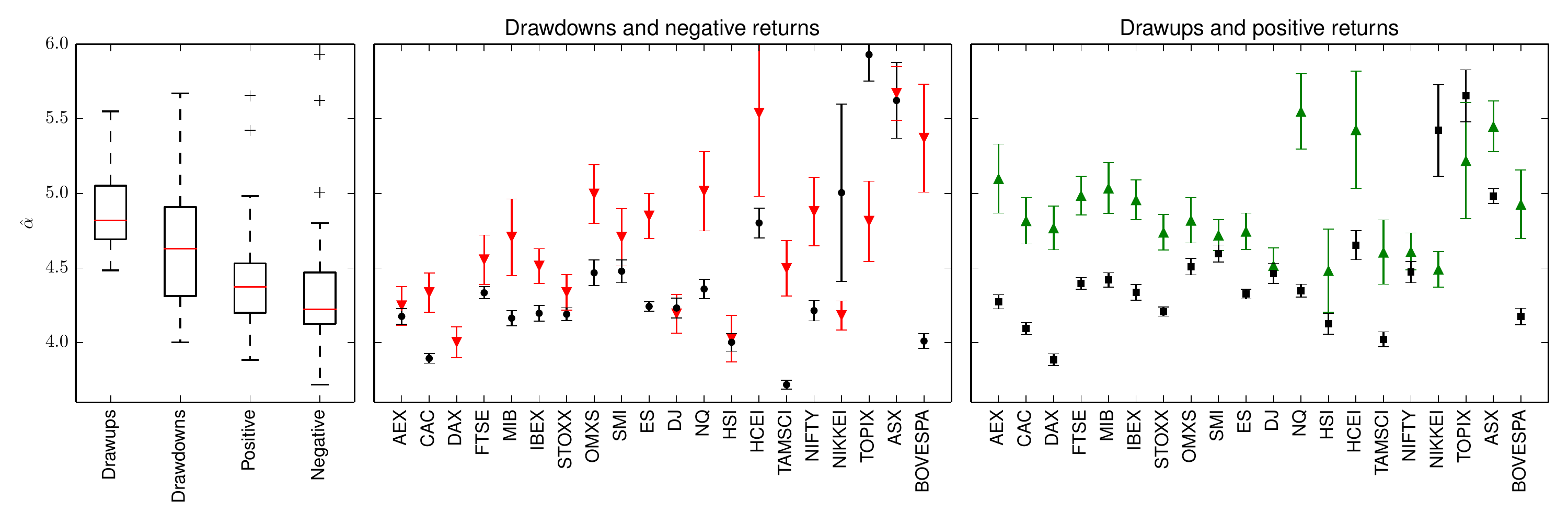}
  \caption{Box plot of estimated exponents $\hat\alpha$ (Table~\ref{tb:fit_dd_norm_returns}); and exponents for drawdowns (red down triangles), drawups (green up triangles), positive (black squares) and negative (black circles) returns for individual contracts.}
\label{fig:boxplot}
\end{figure}

Figure \ref{fig:boxplot} shows that the distributions of both drawups and drawdowns have lighter tails than the distribution of returns. On average, the exponent $\hat\alpha_{du}$ of the power law tail of the distribution of drawups is larger than the corresponding exponent $\hat\alpha_{pos}$ for positive returns by $0.43$. Similarly, the exponent $\hat\alpha_{dd}$ of the power law tail of the distribution of drawdowns is larger than corresponding exponent $\hat\alpha_{neg}$ for negative returns by $0.29$. Overall, the estimated exponents for drawups and drawdowns have values in the interval $4<\hat\alpha<5$ and for log-returns in the interval $3.5<\hat\alpha<4.5$.

These results on the exponents presented in figure \ref{fig:boxplot} are paradoxical, 
since drawdowns cannot, by construction, be smaller than single period returns, as seen from Figure~\ref{fig:ccdf} in which one can observe that the distribution of returns exhibits
a first-order stochastic dominance with respect to the distribution of drawdowns.
As a consequence, the distribution of drawdowns should embody better the extreme loss occurrences. 
Taking at face value that the tail exponents of return distributions are smaller than the tail
exponents of drawdown distributions, as documented in Table~\ref{tb:fit_dd_norm_returns},
would imply that their extrapolations will intersect at some point. 
This would imply that there can be such values of losses that are reached 
with higher probability in a single step rather than in a sequence of steps constituting drawdown 
(and similarly for drawups), which is impossible logically. A first comment is that this
mathematical paradox can never be reached in real life, as this ``intersection'' would occur at unrealistically large values of $x\sim10^6$ (i.e. for returns that are $\sim10^6$ times larger than the volatility of the previous trading day). There is an additional argument to resolve this paradox, which is that 
extrapolating the power law tails assumes that they hold and would hold firmly further for larger events.
But already in our data, we observe significant deviations from the power law tails for the most
extreme events, as shown in figure~\ref{fig:ccdf}, so that a power law approximation becomes highly questionable.

The fact that the distributions of individual returns $r_t^{(\Delta t)}$ have fatter tails than the distribution of normalized returns $r^{norm}$ for drawdowns made of these returns is a signature of
the temporal correlation structure between returns. Moreover, it suggests that drawdowns in the tail of their marginal distribution do not result from the largest individual returns but are constructed 
from a sequence of smaller returns that aggregate transiently either due to 
external or internal market forces.

\begin{figure}[htb!]
  \centering
  \includegraphics[width=\textwidth]{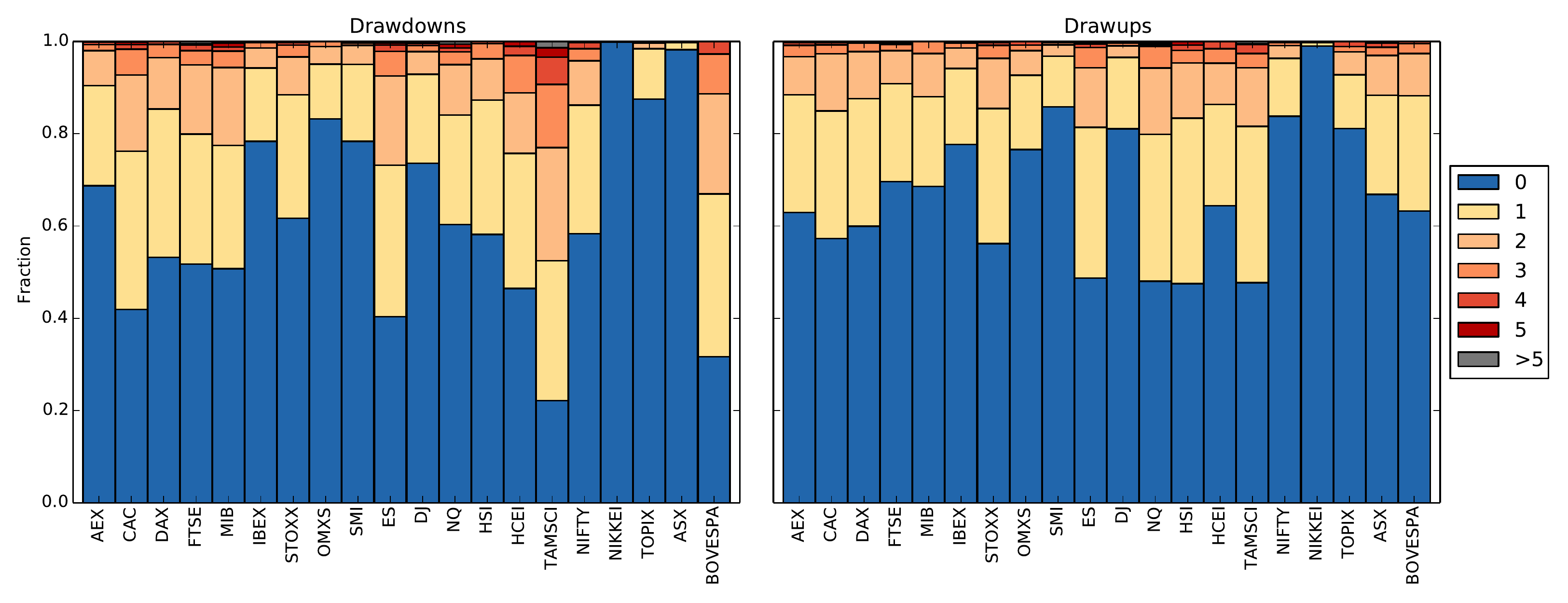}
  \caption{For each analyzed contract, fraction of drawdowns (left) and drawups (right) belonging to the power law tail of their marginal distributions that contain $N=0,1,2,3,4,5$ or $6\le N\le 10$ (color-coded) large returns.}
\label{fig:fraction}
\end{figure}

To test this hypothesis, we calculated the number $N$ of large log-returns (in the power law tail) in each drawdown 
belonging to the power law tail of its normalized return $r^{norm}$ larger than the corresponding $\hat{x}_m$ given in Table~\ref{tb:fit_dd_norm_returns}. For all analyzed drawdowns and drawups, we found that the maximal value of $N$ is 10. Then, for each $N=0,1,\dots,10$, we calculated the number of drawdowns (drawups) that contained $N$ large log-returns. Figure~\ref{fig:fraction} shows that the majority of drawdowns and drawups (more than 50\%) do not contain any log-returns from the tail of the distribution ($N=0$). In particular, the large drawdowns and drawups for NIKKEI and ASX are constructed only from the returns that are smaller than $\hat{x}_m$, i.e. from the body of the distributions.

\begin{figure}[htb!]
  \centering
  \includegraphics[width=0.7\textwidth]{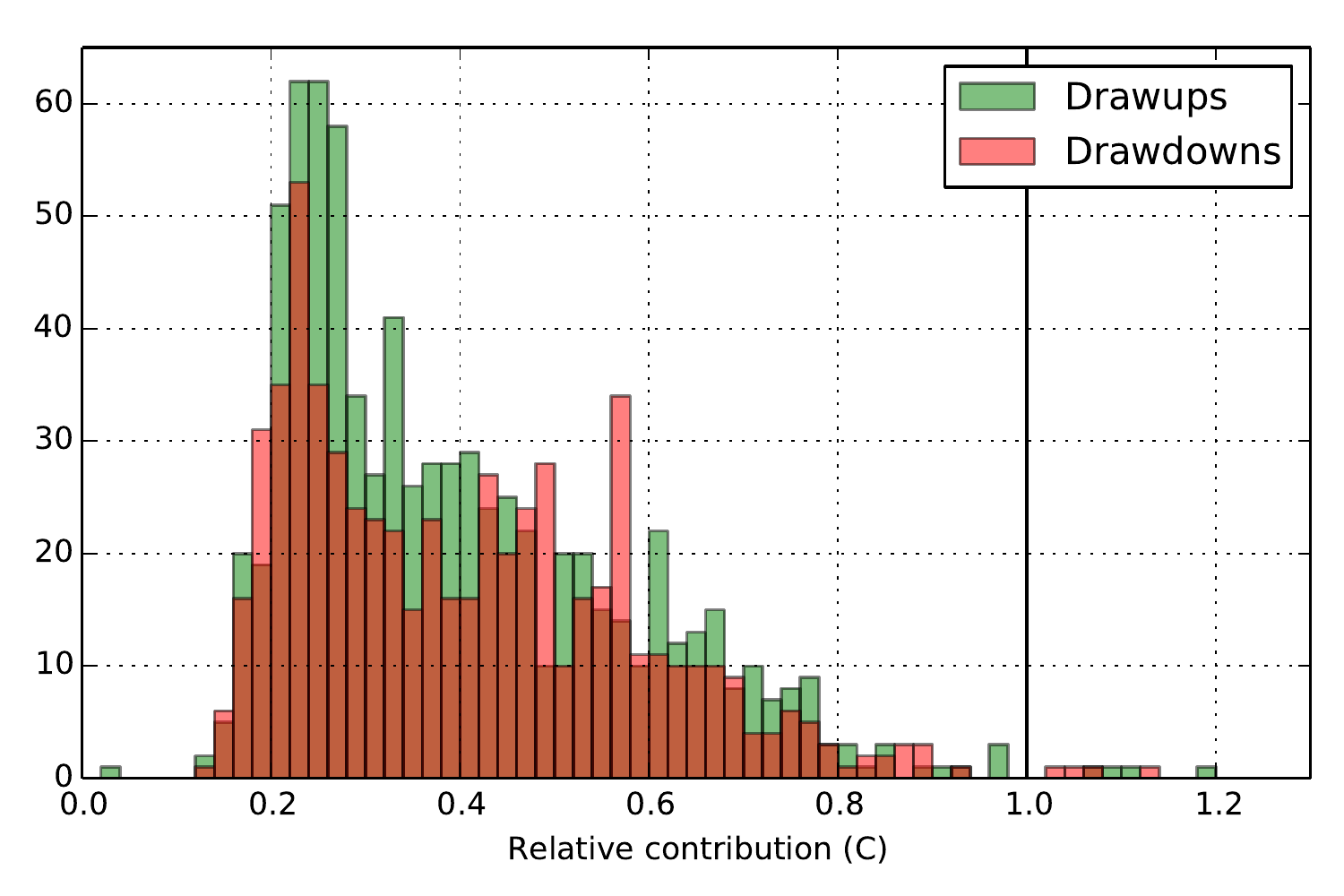}
  \caption{Histogram of the relative contribution ($C$) of large individual log-returns (from the tail of distribution) to the normalized return of large drawdowns (red) and drawups (green) for ES Futures Contract. Vertical line denotes $C=1$ that corresponds to the situation when the drawdown is constructed only from large individual log-returns. }
\label{fig:contribution}
\end{figure}

Moreover, when they are present, the contribution of these large log-returns to the size of the drawdowns (drawups) is not special. In order to illustrate this, for each analyzed contract and each large drawdown/drawup that contain at least one large log-return (i.e. $N\ge1$), we have calculated the relative contribution $C$ of the total sum of the large individual normalized log-returns $\sum r_t^{(\Delta t)}$ to the total normalized return $r^{norm}$ of the event ($C=\sum r_t^{(\Delta t)}/r^{norm}$), which are shown in 
Figure~\ref{fig:contribution} for the E-mini S\&P 500 Futures Contract (codename ES). 
Other contracts give similar qualitative results. One can observe that the distribution of $C$ is skewed towards small values. The mean and median values of $C$ for different contracts vary in the range $0.35-0.45$, indicating that, on average, large individual returns contribute no more than 50\% of the total return
of a drawdown/drawup. Very rarely, one can observe $C\ge1$, which corresponds to the situation when the whole trend is constructed only from individual large returns, and smaller returns of negative sign only provide transient corrections (see the discussion of the drawdown detection method in Section~\ref{sec:dd} and Figure~\ref{fig:drawdown_illustration}).

%===============================================================================
\section{Quantification of the extreme drawdowns}\label{sec:dragonkings}

In the previous section, we have analyzed the tails of the distributions of normalized returns for individual Futures Contracts, which have been found well approximated by power laws over several orders of magnitude. However, some extreme events deviate substantially from the power law tail approximation (see Figure~\ref{fig:ccdf}). The important question discussed in the present section is whether this deviation is statistically significant or not. Statistical significance here is not a mere quantification of goodness-of-fit of distributional characteristic, but addresses the question whether these extreme events are ``outliers'' in the sense of generating mechanisms. If present, we will refer to these special events as ``Dragon Kings'', the notion coined by
one of us \citep{Sornette2009}. To understand what the term means, we first need to explain what are
the implications of power law distributions taken as a reference point.

Recall that power law distributions embody the notion that extreme events are not exceptional events because they belong to the same distribution, which exhibits the remarkable property
of scale invariance: the ratio of the frequencies of two event sizes is proportional to the ratio of the sizes
and independent of the absolute values of the sizes. Only power laws have this property.
In this sense, extreme events are just scaled-up
versions of their  smaller siblings. This is usually interpreted as evidence of the same mechanism
underlying the generation of the whole population, from small to extreme sizes.
As a consequence, because of the common mechanism, these tail events are intrinsically unpredictable
because nothing special takes place before their occurrence that could hint of some special organisation that would put
extreme events apart. In this sense, extreme events in the far right tail of power laws have been
suggested as illustrating the concept of ``black swans'' ~\citep{Taleb2007}. This colourful term is actually
a layman version of the concept of ``unknown unknowns'' introduced by Knight in 1930 
that there exists events that we could not conceive before they happen. Therefore,  strictly in the sense of Knightian
uncertainty, extreme events in the far right tail of power laws are not black swans because we can conceive
of their occurrence since we can quantify their frequencies. Only their specific occurrence time is not predictable.

In contrast, ``Dragon King'' events reveal the existence of mechanisms of self-organization that are not apparent from the (power law) distribution of their smaller siblings. 
Dragon Kings are often associated with a neighborhood of a phase transition, a bifurcation or a tipping point. This distinctive feature (of the approach towards a tipping point) is crucial to learn how to diagnose in advance the symptoms associated with a coming Dragon King (for a more elaborated discussion, we refer to~\citep{Sornette2009,SorouillonDK12}). 

There have already been a number of empirical examples of Dragon Kings documented in the
literature in natural and socio-economic systems, identified statistically as the outliers that occur more frequently that predicted
by the power law distributions calibrated on the rest of the population: stock market crashes~\citep{JohansenSornette2001JofRisk,JohansenSornette1998}; some capitals in the distributions of agglomeration sizes~\citep{PisarenkoSornette2012}; extreme events in the distribution of hydrodynamic turbulent velocity fluctuations, acoustic emissions associated with material failure or epileptic seizures occurring in the strong coupling regime
~\citep{Sornette2009}. Recently, similar ``runaway'' phenomena, that correspond to ``negative'' Dragon Kings, were documented in the distribution of citations~\citep{GolosovskySolomon2012}.

In order to detect Dragon Kings in the distributions, we employ a modified version of the \emph{DK-test} (``Dragon-King test'') proposed by \citet{PisarenkoSornette2012}. The DK-test is based on a statistics that allows one to test quantitatively if the largest $r\ge1$ events in the tail of the empirical distribution belongs to the same distribution 
as the rest of the sample assumed to be a power law \eqref{eq:pl_ccdf}. We start by transforming~\eqref{eq:pl_ccdf} with the following nonlinear mapping:
\begin{equation}\label{eq:pareto_to_exp}
    y=\ln\frac{x}{x_m}.
\end{equation}
If $x$ follows the Pareto distribution~\eqref{eq:pl_ccdf}, then the random variable $y$ is distributed according to the exponential law:
\begin{equation}\label{eq:exp_ccdf}
    F(y)=\mathrm{Pr}\left[Y>x\right] = 1-\exp(-\alpha y),\quad y\ge 0.
\end{equation}
Given independent observations $x_i$ (and thus $y_i)$, we construct a statistical test for the following hypothesis:
\begin{quote}
$H_0$: all observations of the sample are independently generated by the same exponential distribution~\eqref{eq:exp_ccdf}.
\end{quote}
versus its alternative:
\begin{quote}
$H_1$: $r$ observations with the first $r$ ranks ($y_1\ge \dots\ge y_{r-1}\ge y_r$) are generated by a different distribution.
\end{quote}
We consider the differences $\delta y_i=y_{i}-y_{i-1}$ and construct the following auxiliary variables:
\begin{equation}\label{eq:exp_ccdf2}
    z_i=i\cdot \delta y_i=\begin{cases}
       i\cdot(y_{i}-y_{i-1}), \quad &i=1,\dots,N-1;\\
       N\cdot y_N,\quad &i=N.
    \end{cases}
\end{equation}
According to~\citet{PisarenkoSornette2012}, the test statistics
\begin{equation}\label{eq:T}
    T=\frac{z_1+\dots+z_r}{z_{r+1}+\dots+z_N}\cdot\frac{N-r}{r}
\end{equation}
is distributed according to the $f$-distribution with $(2r,2N-2r)$ degrees of freedom. The corresponding p-value for the hypothesis $H_0$ is given by:
\begin{equation}\label{eq:pvalue}
    p(r;y_1,\dots,y_N)=1-F(T, 2r, 2N-2r),
\end{equation}
where $F(t, a, b)$ is the cumulative distribution function (cdf) of the $f$-distribution with $(a,b)$ degrees of freedom. Following~\citet{PisarenkoSornette2012}, we will use threshold the $p_0=0.1$ for rejecting the null hypothesis (for $p<0.1$) and declaring the $r$ largest events as ``outliers''.

We note the existence of a limitation of the original DK-test of~\citet{PisarenkoSornette2012}, 
which appears when the value of the observation with the first rank (or the first few ranks) is extremely large. In this case, this single observation will contribute heavily to the sum in the numerator of expression~\eqref{eq:T}, which will remain large even for large values of $r$. For example, in synthetic samples
of 100 variables distributed according to a pure exponential, adding a single outlier with a size $2\cdot y_1$ (i.e. twice larger than the maximal observed value) is sufficient to 
distort the statistic (\ref{eq:T}) and make it declare that the 10 to 20 largest observations are``outliers'', when using the original formulation of DK-test \citep{PisarenkoSornette2012}. The original DK-test thus does not fail to correctly
identify the presence of Dragon Kings, but it may over-estimate their numbers due to the contamination from 
the existence of a super-large one.

The natural solution to this problem when testing if the $r$-th rank is an outlier is to remove
the $r-1$ largest observations from the set before calculating the DK statistics, In other words, 
when testing if rank $r$ is an outlier, we remove the $r-1$ larger values and make the rank $r$-varialble
the new rank 1. For instance, for $r=2$, we remove the largest observation $r_1$ from the set and then test the null hypothesis versus its alternative that $y_2$ is generated by a different distribution than the rest of the observations ($y_3\ge y_4\ge\dots\ge y_N$). We employ this procedure iteratively starting with $r=1$ in order to find the minimal value of $r$ for which the null hypothesis $H_0$ can no more be rejected.

However, this procedure does not address the lack of power of the DK-test when several large
outliers are present, as occurs for instance in the distributions of drawdowns for the CAC and FTSE contracts shown in Figure~\ref{fig:ccdf}. As an illustration, consider the synthetic sample of 100 exponentially distributed variables $y_1,\dots, y_{100}$, with two additional introduced outliers of sizes respectively  $2\cdot y_1$ and $2.2\cdot y_1$. The application of the iterative procedure described above fails to detect the largest outlier of size $2.2\cdot y_1$. The reason is similar to the one mentioned before: in this case, $y_2$ contributes substantially to the sum in the denominator of expression~\eqref{eq:T} and the overall value of $T$ is not large enough to reject $H_0$.
In other words, the value $2.2\cdot y_1$ is not detected as an outlier in the presence of the value $2\cdot y_1$ together with the other 100 exponentially distributed variables. Removing $2.2\cdot y_1$ and 
applying the DK-test on the variable $2\cdot y_1$ compared with the other remaining
100 exponentially distributed variables $y_1,\dots, y_{100}$ as suggested in our iterative procedure
does diagnose $2\cdot y_1$ as an outlier. This creates a paradox, as the largest value is not
an outlier but the second largest one would be.
 
To overcome all these issues, we propose the following modification of the original DK-test of~\citet{PisarenkoSornette2012}. The $r$ largest observations $y_1,\dots, y_r$
are diagnosed as being ``Dragon-Kings'' or  ``outliers'' if and only if 
the remaining observations $y_{r+1},\dots,y_N$ contains no outliers and if each of 
the variable $y_1,\dots, y_r$ when introduced individually in the remaining set individually can be qualifier as an outlier using the F-test~\eqref{eq:pvalue}. In other words, we will require that the following inequalities are simultaneously satisfied:
\begin{equation}\label{eq:modified_DK}
\begin{cases}
p_1=p(1;y_1,y_{r+1},\dots,y_N)<p_0;\\
p_2=p(1;y_2,y_{r+1},\dots,y_N)<p_0;\\
\dots\\
p_r=p(1;y_{r},y_{r+1},\dots,y_N)<p_0;\\
p_{r+1}=p(1;y_{r+1},y_{r+2},\dots,y_N)\ge p_0.\\
\end{cases}
\end{equation}
This set of conditions (\ref{eq:modified_DK}) captures the logic that, for the first $r$ ranks
to be outliers, the distribution of the other $N-r$ variables is an unperturbed 
exponential law when removing these $r$ outliers, and each of the $r$ outliers
should be individually diagnosed as aberrant.

\begin{table}[p!]
%\vspace{-2ex}
\caption{Characteristics of the extreme drawdowns and drawups for $\Delta t=30$ sec and $\epsilon_0=1$. For each analyzed contract, we report the number of detected ``Dragon-Kings'' ($r$), the timestamp of the beginning of the extreme events (in local time for each Contract --- see Table~\ref{tb:contracts}), their duration $\tau$ in seconds, their size $\Delta P$ in monetary units and the normalized return $r^{norm}$ as well as normalized speed $v^{norm}$. $R_{dur}$ and $R_{vel}$ denote the rank of the events' duration and normalized speed.
The table also presents the p-values given by expression~\eqref{eq:modified_DK} for the modified DK-test for the individual events ($p_k$) and for the remaining observations ($p_{r+1}$). The boldface font indicates p-values for which the null hypothesis $H_0$~\eqref{eq:modified_DK} can be rejected.}
\label{tb:dk_dd}
\begin{center}
%\scriptsize
\renewcommand{\arraystretch}{0.9}
\footnotesize

(a) Drawdowns
%\vspace{0.1cm}

\makebox[\textwidth][c]{
\begin{tabular}{lrrrrrrrrrr}
\toprule
Codename & $r$ &          Timestamp &     $|\Delta P|$ & $|r^{norm}|$ &  $\tau$, sec & $|v^{norm}|$ &  $R_{dur}$ &  $R_{vel}$ &              $p_k$ &             $p_{r+1}$ \\
\midrule
           AEX &           0 &  2005-04-01 17:13:30 &     3.45 &           68.71 &               930 &             0.07 &    1289.0 &        8484 &           0.70 &           0.70 \\
  \textbf{CAC} &  \textbf{3} &  2010-12-27 09:03:00 &   123.00 &          214.99 &               240 &             0.90 &   34381.5 &           1 &  \textbf{0.00} &  \textbf{0.66} \\
               &             &  2005-07-07 11:14:00 &    76.00 &          109.91 &               420 &             0.26 &   11807.0 &          74 &  \textbf{0.04} &  \textbf{0.66} \\
               &             &  2005-07-07 11:24:00 &    61.50 &           89.73 &               180 &             0.50 &   51185.0 &          10 &  \textbf{0.09} &  \textbf{0.66} \\
  \textbf{DAX} &  \textbf{2} &  2010-12-27 09:02:30 &   107.00 &          119.51 &               480 &             0.25 &    7728.0 &         118 &  \textbf{0.03} &  \textbf{0.73} \\
               &             &  2005-07-07 11:14:00 &    86.50 &          112.25 &               390 &             0.29 &   13186.5 &          59 &  \textbf{0.04} &  \textbf{0.73} \\
 \textbf{FTSE} &  \textbf{2} &  2005-07-07 10:14:00 &   113.50 &          176.46 &               270 &             0.65 &   30586.0 &           2 &  \textbf{0.00} &  \textbf{0.96} \\
               &             &  2005-07-07 09:52:30 &    63.50 &           97.73 &               720 &             0.14 &    2250.5 &         796 &  \textbf{0.06} &  \textbf{0.96} \\
           MIB &           0 &  2005-07-07 11:14:30 &   415.00 &           88.45 &               420 &             0.21 &   11749.0 &         117 &           0.12 &           0.93 \\
 \textbf{IBEX} &  \textbf{1} &  2005-07-07 11:14:00 &   206.00 &          147.75 &               390 &             0.38 &   13699.5 &          12 &  \textbf{0.03} &  \textbf{0.47} \\
         STOXX &           0 &  2005-07-07 11:14:00 &    57.00 &           78.21 &               420 &             0.19 &    9811.0 &         160 &           0.25 &           0.32 \\
          OMXS &           0 &  2010-06-04 14:30:00 &    16.00 &           52.25 &               270 &             0.19 &   22339.0 &         143 &           0.85 &           0.36 \\
           SMI &           0 &  2005-07-07 11:25:30 &    99.00 &          121.38 &                90 &             1.35 &   92221.0 &           1 &           0.24 &           0.15 \\
   \textbf{ES} &  \textbf{1} &  2010-05-06 13:41:30 &    59.50 &          132.62 &               240 &             0.55 &   25661.0 &           1 &  \textbf{0.02} &  \textbf{0.39} \\
   \textbf{DJ} &  \textbf{1} &  2010-05-06 13:41:30 &   551.00 &          165.04 &               240 &             0.69 &   23262.0 &           1 &  \textbf{0.03} &  \textbf{0.85} \\
   \textbf{NQ} &  \textbf{1} &  2010-05-06 13:41:30 &   120.00 &          149.19 &               300 &             0.50 &   17303.0 &           1 &  \textbf{0.01} &  \textbf{0.54} \\
           HSI &           0 &  2006-03-07 10:00:30 &   250.00 &           62.50 &               210 &             0.30 &   11541.0 &           4 &           0.22 &           0.84 \\
          HCEI &           0 &  2011-03-15 10:16:30 &   211.00 &           44.33 &               270 &             0.16 &    7525.5 &          65 &           0.93 &           0.31 \\
        TAMSCI &           0 &  2009-09-10 08:53:00 &     4.90 &           41.45 &               210 &             0.20 &   22527.5 &          82 &           0.95 &           0.61 \\
         NIFTY &           0 &  2011-06-20 10:09:30 &    96.10 &           67.75 &               210 &             0.32 &   24678.0 &          12 &           0.16 &           0.79 \\
        NIKKEI &           0 &  2011-03-15 12:38:30 &   240.00 &           27.77 &               330 &             0.08 &    3933.5 &        1561 &           0.53 &           0.98 \\
         TOPIX &           0 &  2007-07-12 14:15:30 &    10.50 &           30.89 &               480 &             0.06 &    1652.0 &        3252 &           0.97 &           0.90 \\
           ASX &           0 &  2009-11-27 10:15:30 &    41.00 &           36.87 &               690 &             0.05 &    2583.5 &       13558 &           0.87 &           0.34 \\
       BOVESPA &           0 &  2011-04-18 10:01:00 &  1310.00 &           50.41 &               150 &             0.34 &   23672.5 &          15 &           0.24 &           0.84 \\
\bottomrule
\end{tabular}
}

\vspace{0.3cm}
 (b) Drawups
\vspace{0.1cm}

\makebox[\textwidth][c]{
\begin{tabular}{lrrrrrrrrrr}
\toprule
Codename &   $r$ &         Timestamp &     $\Delta P$ & $r^{norm}$ &  $\tau$, sec & $v^{norm}$ &  $R_{dur}$ &  $R_{vel}$ &              $p_k$ &             $p_{r+1}$\\
\midrule
             AEX &           0 &  2007-08-17 15:29:00 &    11.05 &           48.71 &               540 &             0.09 &    7197.5 &        4229 &           0.91 &           0.59 \\
    \textbf{CAC} &  \textbf{1} &  2010-12-27 09:07:00 &    90.00 &          157.99 &               240 &             0.66 &   35422.0 &           3 &  \textbf{0.01} &  \textbf{0.19} \\
             DAX &           0 &  2007-08-17 14:13:30 &   136.50 &           52.52 &               180 &             0.29 &   52378.0 &          51 &           0.43 &           0.88 \\
            FTSE &           0 &  2005-07-07 10:18:30 &    50.50 &           79.00 &                30 &             2.63 &  166204.5 &           1 &           0.13 &           0.91 \\
             MIB &           0 &  2011-07-12 10:28:30 &   585.00 &           50.35 &               480 &             0.10 &    9399.5 &        2023 &           0.40 &           0.68 \\
            IBEX &           0 &  2005-07-07 11:36:00 &    72.00 &           51.99 &               240 &             0.22 &   34851.5 &          83 &           0.48 &           0.95 \\
           STOXX &           0 &  2007-12-12 14:58:30 &    69.00 &           59.65 &               300 &             0.20 &   20796.0 &         109 &           0.32 &           0.68 \\
            OMXS &           0 &  2008-01-22 14:19:30 &    39.00 &           43.85 &               240 &             0.18 &   26568.5 &         170 &           0.82 &           0.54 \\
             SMI &           0 &  2005-07-07 11:34:30 &    48.00 &           59.08 &               180 &             0.33 &   50491.0 &          16 &           0.44 &           0.71 \\
              ES &           0 &  2010-05-06 13:45:30 &    37.00 &           83.32 &               180 &             0.46 &   41386.0 &           3 &           0.46 &           0.58 \\
              DJ &           0 &  2007-02-27 14:06:00 &   118.00 &           72.96 &               390 &             0.19 &    8516.0 &         132 &           0.36 &           0.87 \\
     \textbf{NQ} &  \textbf{2} &  2010-05-06 13:46:30 &    76.50 &           96.24 &               120 &             0.80 &   63285.5 &           1 &  \textbf{0.02} &  \textbf{0.55} \\
                 &             &  2010-05-06 13:49:30 &    66.00 &           80.39 &               270 &             0.30 &   21747.5 &          18 &  \textbf{0.07} &  \textbf{0.55} \\
             HSI &           0 &  2005-12-28 09:46:00 &    85.00 &           46.74 &               150 &             0.31 &   18731.5 &           4 &           0.64 &           0.87 \\
            HCEI &           0 &  2007-02-28 10:08:00 &   214.00 &           44.19 &               300 &             0.15 &    6342.0 &         113 &           0.42 &           0.50 \\
 \textbf{TAMSCI} &  \textbf{1} &  2009-09-10 08:46:30 &    14.80 &          127.45 &               390 &             0.33 &    7408.0 &          11 &  \textbf{0.01} &  \textbf{0.80} \\
           NIFTY &           0 &  2008-01-21 14:51:00 &   177.35 &           48.81 &               240 &             0.20 &   20630.0 &          59 &           0.46 &           0.84 \\
          NIKKEI &           0 &  2011-03-15 13:09:30 &   270.00 &           31.14 &               510 &             0.06 &    1223.0 &        5485 &           0.77 &           0.90 \\
           TOPIX &           0 &  2011-03-15 13:14:30 &    23.00 &           35.18 &               210 &             0.17 &   10823.0 &          28 &           0.36 &           0.98 \\
             ASX &           0 &  2008-10-07 14:29:00 &   109.00 &           44.04 &               210 &             0.21 &   32578.5 &          30 &           0.21 &           0.90 \\
         BOVESPA &           0 &  2011-12-01 10:17:30 &  1600.00 &           51.96 &               210 &             0.25 &   15834.0 &          33 &           0.21 &           0.57 \\
\bottomrule
\end{tabular}
}
\end{center}
%\vspace{0.5cm}
\end{table}

Table~\ref{tb:dk_dd} presents the results of the modified DK-test for the distribution of normalized returns $r^{norm}$ of drawdowns and drawups for different contracts. For each test~\eqref{eq:pvalue}, we have sampled the $N=200$ largest events of the distribution tail. The largest number of ``Dragon-Kings'' were found in drawdowns of the CAC contract (3 events), then 2 ``Dragon-Kings'' in the drawdowns of the DAX, and FTSE and in the drawups of the NQ contracts; and 1 ``Dragon-King'' was detected in the drawdowns of the IBEX, ES, DJ and NQ and in the drawups of the CAC and TAMSCI contracts.

Table~\ref{tb:dk_dd} documents that the most famous intraday price swing, the so-called ``flash crash'' of May 6, 2010 (13:41:30 EST), can be diagnosed as a ``Dragon King'' event for all three US E-mini futures contracts: ES, DJ and NQ. Over the time of 4 minutes (5 for NQ), the price of these index futures contracts dropped by $r^{norm}=132.62$, $165.04$ and $149.19$ respectively, which means that the price drop was more than a 130-sigma event (i.e., 132--165 times larger than the volatility of 30-second returns). As shown by~\citet{FilimonovSornette2012_Reflexivity}, the dynamics of the high-frequency mid-quote price during this ``flash-crash'' exhibited a unique pattern, indicating an extreme degree of self-excitation during that event. As reported, slightly before and during the price drop, the system became critical, being essentially driven by the internal feedback mechanisms rather than the external information flow.

Two consecutive drawups occurring after the ``flash crash'' of NQ (13:46:00 and 13:49:30 EST with $r^{norm}=96.24$ and $80.39$) can be also claimed to be ``outlier'' of the respective probability distribution. The ``flash crash'' started on E-mini S\&P 500 futures contracts~\citep{FlashCrash2010_report}, and then almost instantly propagated to the constituting stocks of the index as a result of the arbitrage between ETF and Futures and between ETF and the underlying assets~\citep{BenDavid2011}. The ``Flash crash'' of May 6, 2010 was attributed to the activity of high-frequency traders according to the joint SEC and CFTC report~\citep{FlashCrash2010_report}. These high-frequency traders did not trigger the crash but contributed significantly to the market volatility and extraordinarily amplified the initial price drop.

While being the best known, the ``flash crash'' of May 6, 2010 is not the largest one in relative values in the data analysed here. A remarkable drawdown of $r^{norm}=214.99$ (``215-sigma event'') was experienced in the futures on the CAC index on December 27, 2010 at 09:03:00 CET. This ``mini flash crash'' at the opening of that trading day initially started on the DAX index
and quickly propagated to other other European markets through cross-market arbitrage, many of which experienced large drawdowns. However, only for the CAC and DAX futures contracts can this event be quantified as a ``Dragon King'' according to our modified DK-test (\ref{eq:modified_DK}). The CAC futures contract recovered from the ``mini flash crash'' via
an extreme drawup of size $r^{norm}=157.99$, which is also qualified as a dragon-king (the hypothesis $H_0$ that this event belongs to the overall distribution can be rejected).

These two ``flash crashes'' in US markets on May 6 and in Europe on December 27, 2010 represent endogenous events, where were not generated by external news but by a self-exciting activity of the market participants. But not all dragon-king events are generated internally, as illustrated by  
the plunge of European markets in response to the 7 July 2005 London bombings, when a series of coordinated suicide attacks targeted the civilian public transportation system in central London during the morning rush hour. At that time, almost all analyzed contracts experienced an outstanding drop and, for the CAC, DAX, FTSE and IBEX contracts, these price fluctuations can be quantified as ``Dragon Kings'' of their respective drawdown distributions.

In general, the Asian, Australian and South American contracts do not exhibit any
outstanding price fluctuations that can be quantified as ``Dragon King'', with a single exception. On the opening of the Taiwanese markets on September 10, 2009, the price of TAMSCI futures contracts experienced a drawup with $r^{norm}=127.45$. After the drawup of the CAC contract following the ``mini flash crash''
described above, this is the largest normalized return over all observed drawups for all analyzed contracts. This extraordinary drawup of the market resulted from the announcements of an agreement between Taiwan and mainland China to allow mainland investors to buy stocks in Taiwan.

\begin{table}[ht!]
%\vspace{-2ex}
\caption{Characteristics of the ``Dragon King'' events in terms of (a) normalized speed $v^{norm}$ and (b) duration $\tau$. For all analyzed contracts, only those set of events, for which hypothesis $H_0$ can be rejected, are presented. $R_{ret}$, $R_{dur}$ and $R_{vel}$ denote the rank of the normalized return, duration and normalized speed of the corresponding event.}
\label{tb:dk_dd_dur_speed}
\begin{center}
%\scriptsize
\renewcommand{\arraystretch}{0.9}
\footnotesize

(a) Normalized velocity, $v^{norm}$
\makebox[\textwidth][c]{
\begin{tabular}{lrrrrrrrrrr}
\toprule
Codename & $r$ &          Timestamp &     $|\Delta P|$ & $|r^{norm}|$ &  $\tau$, sec & $|v^{norm}|$ &  $R_{ret}$ &  $R_{dur}$ &              $p_r$ &             $p_{r+1}$ \\
\midrule
\multicolumn{11}{c}{Drawdowns}\\
\midrule
 \textbf{FTSE} &  \textbf{1} &  2011-09-02 13:30:00 &  81.50 &           34.90 &                30 &             1.16 &         16 &    165956 &  \textbf{0.09} &  \textbf{0.67} \\
\midrule
\multicolumn{11}{c}{Drawups}\\
\midrule
   \textbf{FTSE} &  \textbf{1} &  2005-07-07 10:18:30 &   50.50 &           79.00 &                30 &             2.63 &          1 &  166204.5 &  \textbf{0.00} &  \textbf{0.79} \\
     \textbf{ES} &  \textbf{1} &  2007-09-18 13:14:30 &   25.50 &           71.18 &                60 &             1.19 &          2 &   99839.5 &  \textbf{0.03} &  \textbf{0.95} \\
 \textbf{NIKKEI} &  \textbf{1} &  2005-12-08 12:58:30 &  100.00 &           16.70 &                30 &             0.56 &         10 &   59106.0 &  \textbf{0.06} &  \textbf{0.62} \\
\bottomrule
\end{tabular}
}

\vspace{0.3cm}
(b) Duration, $\tau$
\vspace{0.1cm}

\makebox[\textwidth][c]{
\begin{tabular}{lrrrrrrrrrr}
\toprule
Codename & $r$ &          Timestamp &     $|\Delta P|$ & $|r^{norm}|$ &  $\tau$, sec & $|v^{norm}|$ &  $R_{ret}$ &  $R_{vel}$ &              $p_r$ &             $p_{r+1}$ \\
\midrule
\multicolumn{11}{c}{Drawdowns}\\
\midrule
     \textbf{ES} &  \textbf{1} &  2008-07-04 08:42:30 &   7.25 &           14.03 &              5670 &             0.00 &        436 &      135255 &  \textbf{0.10} &  \textbf{0.43} \\
    \textbf{HSI} &  \textbf{1} &  2006-03-08 10:35:00 &  66.00 &            4.70 &              3000 &             0.00 &       8871 &       47291 &  \textbf{0.05} &  \textbf{0.82} \\
 \textbf{TAMSCI} &  \textbf{1} &  2005-02-16 12:25:30 &   0.70 &            1.71 &              4770 &             0.00 &      66109 &      100446 &  \textbf{0.02} &  \textbf{0.93} \\
\midrule
\multicolumn{11}{c}{Drawups}\\
\midrule
 \textbf{TAMSCI} &  \textbf{1} &  2005-02-16 09:41:00 &  2.80 &            6.86 &              5640 &             0.00 &       5936 &      100478 &  \textbf{0.03} &  \textbf{0.95} \\
\bottomrule
\end{tabular}
}
\end{center}
%\vspace{0.5cm}
\end{table}

Table~\ref{tb:dk_dd} shows that some of the reported extreme (in terms of normalized return) drawdowns and drawups are also the fastest. This is the case for the drawdowns observed for CAC, SMI, ES, DJ and NQ and for the drawups for  FTSE and NQ. However, the normalized speeds of these events are not exceptional. Moreover, from the point of view of the speed statistics, these events cannot be quantified as ``Dragon Kings''. Table~\ref{tb:dk_dd_dur_speed} (a) reports the ``outliers'' that our modified DK-test
detects in the distributions of the normalized speeds. None of the reported ``Dragon Kings'' of the normalized returns (Table~\ref{tb:dk_dd}) is also exceptional in terms of speed. The only common event in these two tables --- the drawup for the FTSE contract on 2005-07-07 10:18:30 GMT --- cannot be quantified as a ``Dragon King'' with respect to its normalized return $r^{norm}$ (this is vividly illustrated in Figure~\ref{fig:ccdf}). It is also interesting to observe that, in general, the power law fit of the distribution of normalized speeds is more robust than the distribution of normalized returns (Figure~\ref{fig:ccdf}).

Finally, Table~\ref{tb:dk_dd_dur_speed} (b) reports the ``Dragon Kings'' detected in the 
distributions of duration $\tau$ and shows that all these events are not even close to be the largest or the fastest. This raises naturally the issue of the dependence between these different measures.
For this, we will discuss the so-called tail dependence (dependence of extreme values) later in Section~\ref{sec:dependence}.

%===============================================================================
\section{Aggregated distributions}\label{sec:distrib_aggragated}

In the previous section, we have analyzed the distribution of normalized returns for individual contracts. As discussed above, we are considering normalized characteristics of drawups and drawdowns (such as returns $r^{norm}$ and speeds $v^{norm}$), which allows us to compare directly these values for events of different time periods and different assets. Moreover, we can aggregate all values from different contracts, to check if our conclusions are robust with respect to sampling.

\begin{figure}[h!]
  \centering
  \includegraphics[width=\textwidth]{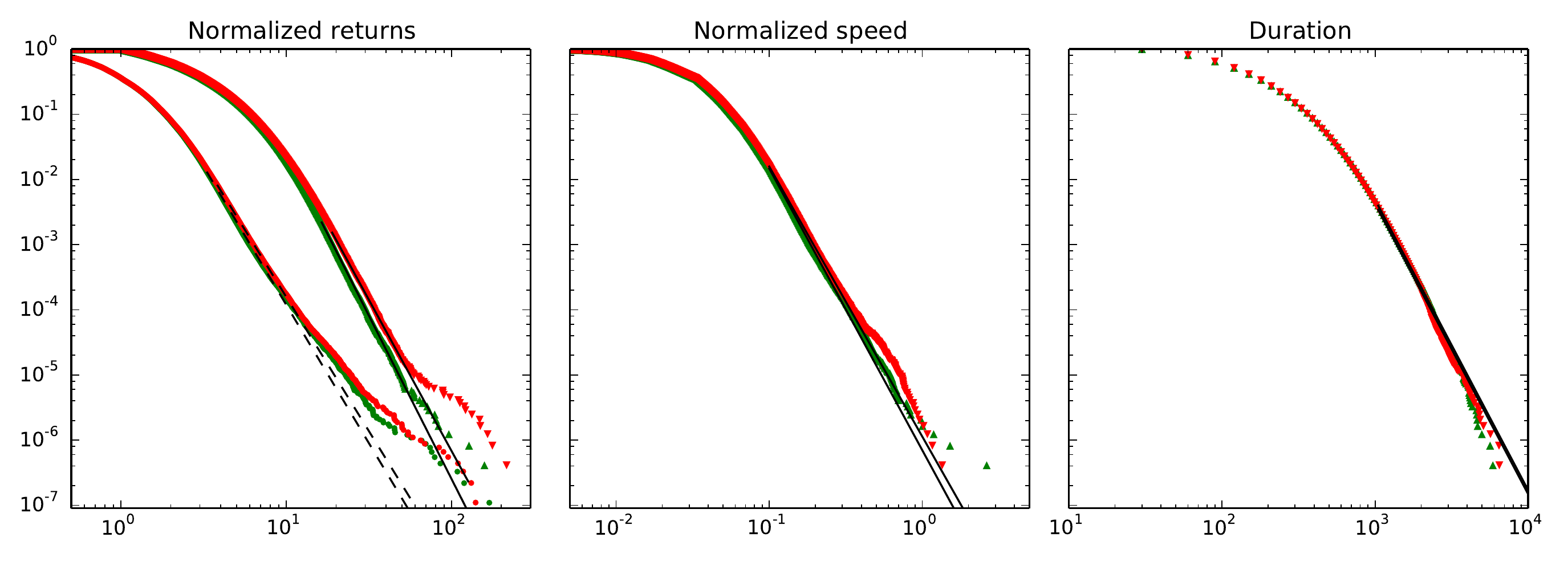}
  \caption{Complementary cumulative distribution function (ccdf) for the (i) aggregated normalized returns, (ii) aggregated normalized speeds and (iii) aggregated durations of drawdowns (red down triangles) and drawups (green up triangles) for $\epsilon_0=1$ and $\Delta t=30$ sec. Black straight lines correspond to power law fits of the tails of distributions of drawdowns (see Table~\ref{tb:fit_aggregated}). Red and green dots on the plot (i) correspond to distributions of the aggregated normalized log-returns~\eqref{eq:ret} over the time scale $\Delta t=30$ sec. The dashed black lines depict the power law fits of the tails of these distributions (see Table~\ref{tb:fit_aggregated}).}
\label{fig:ccdf_aggregated}
\end{figure}

Figure~\ref{fig:ccdf_aggregated} presents the corresponding aggregated empirical distribution functions for the normalized returns, speeds and durations, and the power law approximations of the tails of these distributions. Table~\ref{tb:fit_aggregated} lists the parameters of the power law fits~\eqref{eq:pl_ccdf}.

\begin{table}[ht!]
\caption{Estimates of the lower boundary $\hat x_m$ and exponent $\hat \alpha$ of the power law fits~\eqref{eq:pl_ccdf} of the distributions of (i) normalized returns $r^{norm}$, (ii) normalized speeds $v^{norm}$ and (iii) durations $\tau$ for normalized drawdown (DD) and drawup (DU), as well as for normalized log-returns~\eqref{eq:ret}: positive (RPos) and negative (Rneg). 
The number of observations qualified in the power law tail ($N_{x\ge \hat x_m}$), log-likelihood ratio ($\mathcal{R}$) and p-values for the significance of likelihood ratio test are also given.}
\label{tb:fit_aggregated}
\begin{center}
%\scriptsize
\renewcommand{\arraystretch}{0.9}
\footnotesize
\begin{tabular}{ccrcrrr}
\toprule
                        Characteristic & Event type &     $\hat x_m$ &            $\hat\alpha$ &     $N_{x\ge \hat x_m}$ \\
\midrule
 \multirow{4}{*}{$r^{norm}$} &    DD &    17.62 &  4.64 $\pm$ 0.06 &    5197  \\
                             &    DU &    16.03 &  5.01 $\pm$ 0.07 &    5943  \\
                             &  RNeg &     3.55 &  4.12 $\pm$ 0.01 &  101512  \\
                             &  RPos &     3.03 &  4.25 $\pm$ 0.01 &  172093  \\
\midrule
 \multirow{2}{*}{$v^{norm}$} &    DD &     0.10 &  4.14 $\pm$ 0.02 &   39677  \\
                             &    DU &     0.10 &  4.32 $\pm$ 0.02 &   35911  \\
\midrule
     \multirow{2}{*}{$\tau$} &    DD &   960.00 &  4.46 $\pm$ 0.04 &   12322  \\
                             &    DU &  1020.00 &  4.46 $\pm$ 0.04 &   10871  \\
\bottomrule
\end{tabular}
\end{center}
\end{table}

These results validate our previous findings. First, fits of the distribution of drawdowns/drawups are much more robust than for individual log-returns. 
Second, the reported exponents of the power law tails for the aggregated distributions 
lie in the same range of values that we reported for individual contracts (see Figure~\ref{fig:boxplot}). 
Third, the estimated exponent $\hat\alpha$ for drawdowns is significantly larger than 
the exponent  $\hat\alpha$ for drawups (the difference is larger than 6 standard errors). 

Figure~\ref{fig:ccdf_aggregated} and Table~\ref{tb:fit_aggregated} show that 
the power law approximation of the normalized speeds of drawdowns and drawups is almost perfect and holds for more than 5 orders of magnitude in the vertical axis. In contrast, the fits of the tails of the distributions of durations are relatively poor. In particular, the hypothesis that the distribution of drawup durations is a power law can be rejected in favor of the stretched exponential distribution family using the nested Wilks' test~\citep{Sornette2005}.

An important observation from Figure~\ref{fig:ccdf_aggregated} is that the extreme events 
of individual distributions (Figure~\ref{fig:ccdf}) also branch off the aggregated distribution. One can clearly see that up to ten of the largest drawdowns and drawups deviate substantially from the power law fit of the tail. Interestingly, the original~\citep{PisarenkoSornette2012} and the modified~(Section~\ref{sec:dragonkings}) DK-tests give contradictory and confusing results.
However,  ``the absence of evidence 
is not the evidence of absence'', which summarises the fallacy of the argumentum ad ignorantiam.
In other words, we argue that the failure to diagnostic the largest drawdowns as outliers reflects the
lack of power of these tests. As discussed above, the 
15--20 extremes events in the tail contribute substantially to the numerator in
expression \eqref{eq:T} and lead to a spurious identification of up to 400 outliers in the tail
(the  $H_0$ hypothesis is rejected for values of ranks $r$ up to $r=400$). 
On the another hand, the modified DK-test~\eqref{eq:modified_DK} sins at the other
extreme by being too conservative and fails to reject $H_0$ for any $r$, 
because it requires the simultaneous rejection of $H_0$ for $y_{r},y_{r+1},\dots,y_N$ and 
the acceptance of $H_0$ for  $y_{r+1},\dots,y_N$. This typically does not occur
when outliers are not a few far-standing events, but are organised
with a continuous and smooth deviation of the tail as in Figure~\ref{fig:ccdf_aggregated}.

To address this issue and test for the presence of a change of regime
in the tail of the distribution, we employ the parametric U-test~\citep{PisarenkoSornette2012}, which tests deviation of the tail with respect to the fitted power law distribution rather than with respect to the rest of sample as in the DK-test. We present a slightly modified description of the U-test from that found in 
\citep{PisarenkoSornette2012} and provide a closed-form solution~\eqref{eq:alpha_exp} of the maximum likelihood estimation of the power law exponent.

We select the lower threshold $x_m$ for the calibration of the power law in the distribution tail and apply  the same nonlinear transformation~\eqref{eq:pareto_to_exp} as for the DK-test. Then, by visual inspection, we determine a candidate for the rank $r$ such that observations smaller than $y_{r+1}$ (i.e.,
of rank larger than $r$) are distributed according to the exponential distribution~\eqref{eq:exp_ccdf} and the total number of outliers is not larger than $r$. 

The exponent $\alpha$ of the exponential distribution \eqref{eq:exp_ccdf} can be estimated with the Maximum Likelihood method applied to the subsample $y_{r+1},\dots,y_N$, where the likelihood with right-censored observations is given by
\begin{equation}\label{eq:lik}
	\mathcal{L}(\alpha|y_{r+1},\dots,y_N)=\big[1-F(y_{r+1})\big]^r\prod_{k=r+1}^Nf(y_k),
\end{equation}
where $F(y)$ is the cdf of the exponential distribution~\eqref{eq:exp_ccdf} and $f(x)=\alpha\exp(-\alpha x)$ is the corresponding pdf. 
%Note that the likelihood~\eqref{eq:lik} is different from standard form of likelihood for \todo{censored data}. Rather than normalizing theoretical density in the \todo{censored range} $[0,y_{r+1}]$, we directly use information that $r$ observations in our sample are larger than $y_{r+1}$. 
The exponent $\alpha$ can then be estimated by maximizing $\log\mathcal{L}$. In  the case of an exponential distribution~\eqref{eq:exp_ccdf}, this yields the closed form expression
\begin{equation}\label{eq:alpha_exp}
	\hat\alpha=(N-r)\cdot\left[ry_{r+1}+\sum_{k=r+1}^Ny_k\right]^{-1}
\end{equation}
The p-values that the $k$ smallest ranks deviate from the null hypothesis of the exponential distribution can be then obtained from the following equation (see derivations in~\citet{PisarenkoSornette2012}):
\begin{equation}\label{eq:pvalue_U}
	p_k=1-\mathcal{B}\big(F(y_k);n-k+1, k\big),\quad k=1,\dots, r,
\end{equation}
where $\mathcal{B}(y; a, b)$ is the normalized incomplete beta-function, and the exponent $\alpha$ for the probability distribution~\eqref{eq:exp_ccdf} is taken to be equal to the MLE (\ref{eq:alpha_exp}). 
The event $y_k$ for which $p_k<p_0=0.1$ can be then diagnosed as an outlier with respect to the fitted exponential distribution of the tail. This corresponds to the original event $x_k$ being a ``Dragon-King'' with respect to the fitted power law distribution.

\begin{table}[tb!]
\caption{Detected outliers for the aggregated distribution of normalized returns of (a) drawdowns and (b) drawups. Timestamps are given in local time. Individual p-values for each event are calculated using the U-test~\eqref{eq:pvalue_U}. The last column of the tables indicates if an event is the largest in the individual distribution (E), or if it was also quantified as a ``Dragon-King'' in the corresponding individual contract (DK) --- see Table~\ref{tb:dk_dd}.}
\label{tb:dk_aggregated}
\renewcommand{\arraystretch}{0.9}
\footnotesize

\makebox[\textwidth][c]{

\begin{minipage}[t]{0.53\textwidth}
\begin{center}
(a) Drawdowns

\begin{tabular}{lcrcc}
\toprule
 Codename &            Timestamp & $|r^{norm}|$ &       $p$  & \\
\midrule
   CAC &  2010-12-27 09:03:00 &          214.99 &  0.04 &  DK \\
  FTSE &  2005-07-07 10:14:00 &          176.46 &  0.01 &  DK \\
    DJ &  2010-05-06 13:41:30 &          165.04 &  0.00 &  DK \\
    NQ &  2010-05-06 13:41:30 &          149.19 &  0.00 &  DK \\
  IBEX &  2005-07-07 11:14:00 &          147.75 &  0.00 &  DK \\
    ES &  2010-05-06 13:41:30 &          132.62 &  0.00 &  DK \\
   SMI &  2005-07-07 11:25:30 &          121.38 &  0.00 &   E \\
   DAX &  2010-12-27 09:02:30 &          119.51 &  0.00 &  DK \\
   DAX &  2005-07-07 11:14:00 &          112.25 &  0.00 &  DK \\
   CAC &  2005-07-07 11:14:00 &          109.91 &  0.00 &  DK \\
  FTSE &  2005-07-07 09:52:30 &           97.73 &  0.00 &  DK \\
   CAC &  2005-07-07 11:24:00 &           89.73 &  0.00 &  DK \\
   SMI &  2005-07-07 11:14:30 &           88.45 &  0.00 &     \\
   MIB &  2005-07-07 11:14:30 &           88.45 &  0.00 &   E \\
 STOXX &  2005-07-07 11:14:00 &           78.21 &  0.00 &   E \\
    DJ &  2007-02-27 13:51:00 &           72.73 &  0.00 &     \\
    DJ &  2008-09-29 12:42:00 &           70.16 &  0.00 &     \\
   AEX &  2005-04-01 17:13:30 &           68.71 &  0.01 &   E \\
 NIFTY &  2011-06-20 10:09:30 &           67.75 &  0.00 &   E \\
  IBEX &  2007-04-24 10:46:00 &           65.28 &  0.01 &     \\
   AEX &  2010-12-27 09:03:30 &           63.57 &  0.02 &     \\
    ES &  2008-09-29 12:42:00 &           63.45 &  0.01 &     \\
   HSI &  2006-03-07 10:00:30 &           62.50 &  0.02 &   E \\
   \bottomrule
\end{tabular}
\end{center}
\end{minipage}~~\begin{minipage}[t]{0.53\textwidth}
\begin{center}
(b) Drawups

\begin{tabular}{lcrcc}
\toprule
 Codename &            Timestamp & $r^{norm}$ &       $p$ & \\
\midrule
    CAC &  2010-12-27 09:07:00 &          157.99 &  0.06 &  DK \\
 TAMSCI &  2009-09-10 08:46:30 &          127.45 &  0.01 &  DK \\
     NQ &  2010-05-06 13:46:30 &           96.24 &  0.04 &  DK \\
     ES &  2010-05-06 13:45:30 &           83.32 &  0.07 &   E \\
     NQ &  2010-05-06 13:49:30 &           80.39 &  0.04 &  DK \\
   FTSE &  2005-07-07 10:18:30 &           79.00 &  0.02 &   E \\
     DJ &  2007-02-27 14:06:00 &           72.96 &  0.03 &   E \\
     ES &  2007-09-18 13:14:30 &           71.18 &  0.02 &     \\
    CAC &  2007-12-12 14:58:30 &           66.59 &  0.05 &     \\
     ES &  2007-02-27 14:06:30 &           64.01 &  0.06 &     \\
\bottomrule
\end{tabular}
\end{center}
\end{minipage}
}

\end{table}

Table~\ref{tb:dk_aggregated} presents results of the application of the U-test to the aggregated distribution of normalized returns $r^{norm}$ of drawdowns, where the threshold $x_m$ is selected by using the Kolmogorov-Smirnov test (Table~\ref{tb:fit_aggregated}). The first important observation is that all ``Dragon-King'' events that were detected for individual contracts (Table~\ref{tb:dk_dd}) are also quantified as belonging to a different regime than the power law for the aggregated distribution. This supports the findings of Section~\ref{sec:dragonkings}.

However, not all extremes (events with rank 1) of the individual contracts are present n the upper tail of the aggregated distribution. For drawdowns, the extreme tail of the aggregated distribution contains mostly events from  European and US markets. For example, the extreme drawdowns for OMXS, HCEI, TAMSCI, NIKKEI, TOPIX, ASX and BOVESPA are not qualified as an outliers (neither they were reported as ``Dragon-Kings'' at the individual level). For drawups, the number of detected outliers is much smaller than for drawdowns.

On the contrary, several events classified as outliers at the aggregate level were not reported as individual ``Dragon-Kings'' using the (conservative) modified DK-test. For example, we were unable to reject the null hypothesis $H_0$ for the two largest events occurring in the SMI futures contracts (Figure~\ref{fig:ccdf}). For $r=2$, we report the following p-values: $p_1=0.03$, $p_2=0.13$ and $p_3=0.47$; for $r=3$, we obtain p-values: $p_1=0.01$, $p_2=0.05$, $p_3=0.47$ and $p_4=0.99$. In both cases, one inequality of the system~\eqref{eq:modified_DK} does not hold. This results from the fact that 
the second event $x_2$ is not sufficiently larger than $x_3$, which leads
to the absence of rejection of the null (no dragon-kings) for $r=2$ ($p_2=0.13>0.1$). And the third event $x_3$ only slightly deviates from the tail ($p_3=0.47$ for $r=3$), which leads to the 
absence of rejection of the null for larger $r$ values.

Finally, all events detected as outliers of the aggregate distribution (Table~\ref{tb:dk_aggregated}) can be detected in their corresponding individual distributions using the U-test. However, being dependent on the 
calibration of the exponent of the power law, the U-test is subjected to estimation errors that need to be accounted properly. The nonparametric DK-test is free from this drawback, at the
cost of having more limited power. In general, as with any statistical testing problem, it is always a good practice to consider several different tests to confirm the conclusions. 

%===============================================================================
\section{Tail dependence characteristics of extreme drawdowns}\label{sec:dependence}

Are extreme drawdowns (drawups) associated with the largest speed and/or the longer durations?
Clarifying the interdependence between size, speed and duration of extreme
drawdown (drawups) is important to better understand their generating mechanism.
In previous sections, we have already commented that  events that are extreme with respect to one 
characteristic may not be extreme with respect to another (see Tables~\ref{tb:dk_dd} and~\ref{tb:dk_dd_dur_speed}). The largest (with respect to $r^{norm}$) drawdowns and drawups are often not the fastest, and by far not the longest events in the population. The occurrence of an extreme normalized speed $v^{norm}$ does not ensure that the event will have an extreme size. Moreover, the 
longest drawdowns and drawups typically have relatively small returns. 

Our goal here is to quantify the mutual interdependence between size, speed and duration.
Generally, the complete information about the dependence between two random variables $X$ and $Y$ is contained in their copula structure. Here, we consider a simpler metric, the 
tail dependence, which is defined at the probability of observing a very large value
of one variable conditional on the occurrence of a very large value of the other variable
\citep{Sornette_Risks2005}:
\begin{equation}\label{eq:tail_dependence}
	\lambda=\lim_{u\to1}\lambda_u=
	\lim_{u\to1}\mbox{Pr}[X>F_X^{-1}(u)|Y>F_Y^{-1}(u)],
\end{equation}
where $F_X(\cdot)$ and $F_Y(\cdot)$ are the marginal cumulative distributional functions of $X$ and $Y$. In practice, it is difficult to work with the asymptotic tail dependence $\lambda$, which is defined in the 
empirically unattainable limit $u \to 1$. We will thus consider $\lambda_u$ for fixed value of probability $u\lesssim1$ and document the behaviour of $\lambda_u$ as $u$ approaches $1$ from below.

\begin{figure}[t!]
  \centering
  \includegraphics[width=0.7\textwidth]{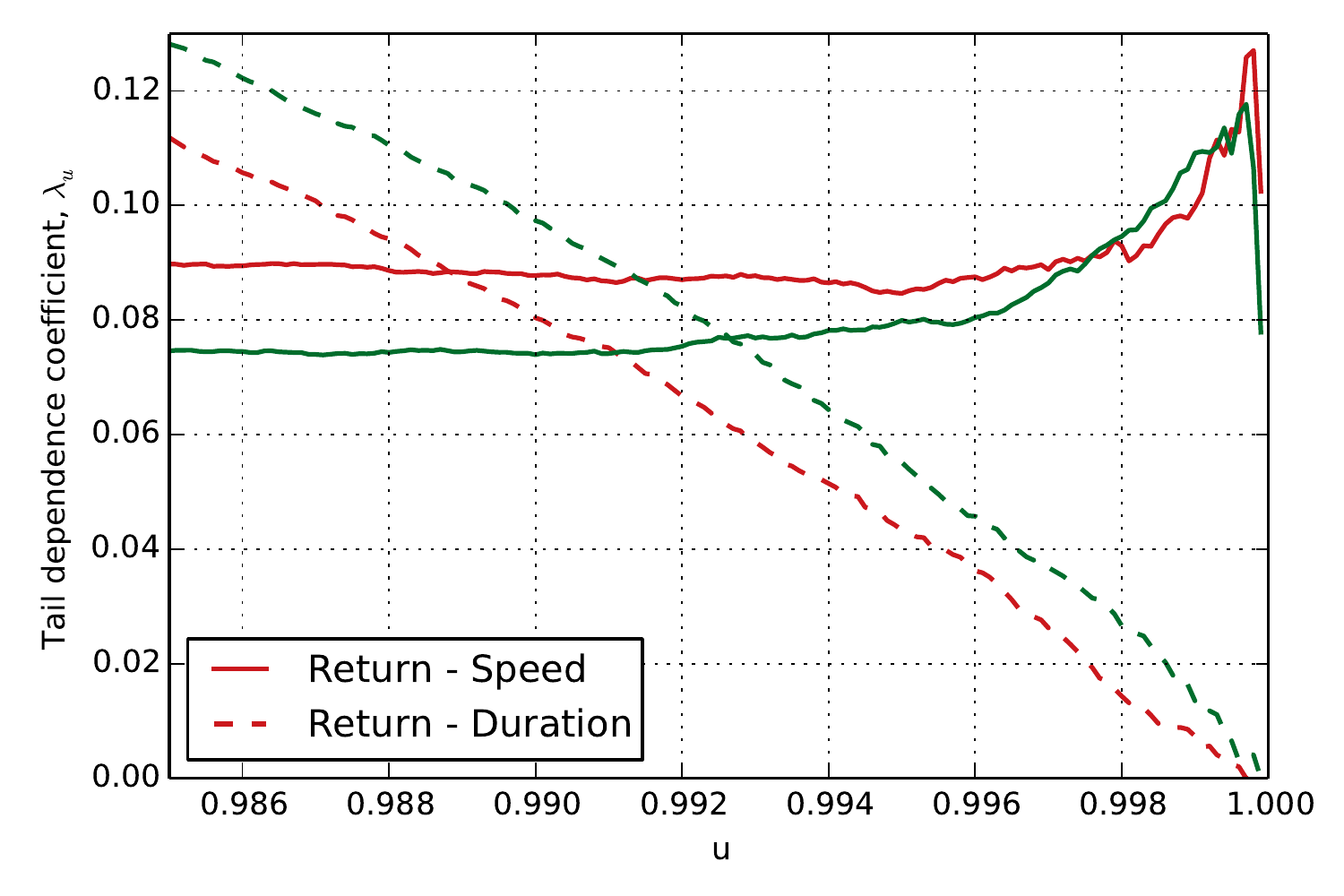}
  \caption{Tail dependence coefficients $\lambda_u$ for normalized returns $r^{norm}$ and normalized speeds $v^{norm}$ (solid lines) and for normalized returns $r^{norm}$ and durations $\tau$ (dashed lines) of the aggregated drawdowns (red) and drawups (green) for different contracts and different probabilities $u$.}
\label{fig:td_aggregated}
\end{figure}

Figure~\ref{fig:td_aggregated} shows the non-asymptotic tail dependence coefficients
$\lambda_u$ of (i) $X=r^{norm}$ and $Y=v^{norm}$ and (ii) $X=r^{norm}$ and $Y=\tau$ for the aggregated probability distributions $F(r^{norm}, v^{norm})$ and $F(r^{norm}, \tau)$ (marginal distributions are presented in Figure~\ref{fig:ccdf_aggregated}). In other words, figure~\ref{fig:td_aggregated} quantifies
the probability that the observed drawdown is large, conditional on it being (i) fast or (ii) long.  
One can observe that $\lambda_u$ of the normalized returns conditional on the durations 
decreases monotonously with $u$ and converges to zero as $u\to1$. This indicates
an absence of dependence of the extreme values of size and durations. In other words, the longest drawdowns and drawups do not belong to the highest quantiles in term of sizes.

\begin{figure}[ht!]
  \centering
  \includegraphics[width=\textwidth]{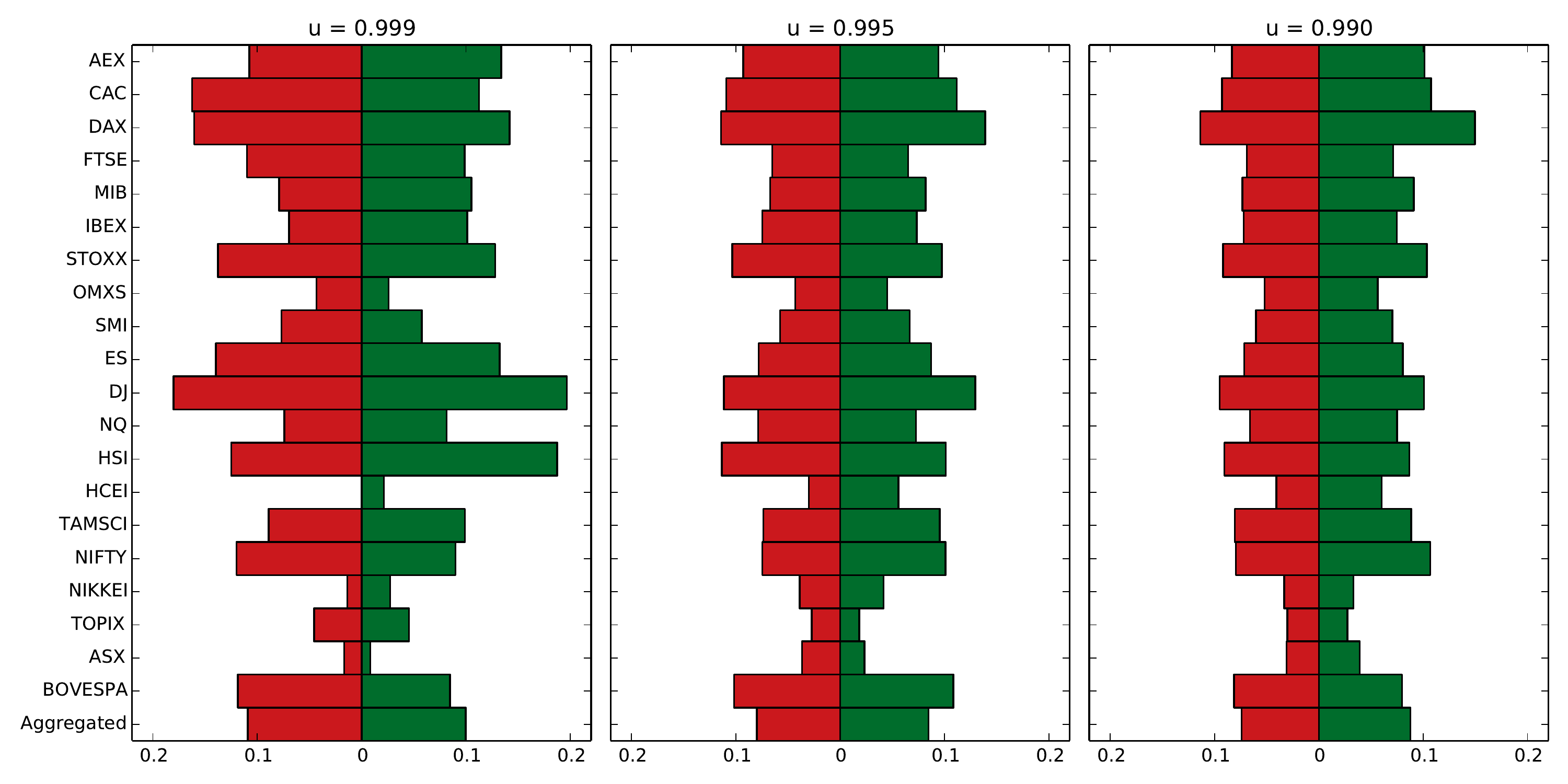}
  \caption{Tail dependence coefficient $\lambda_u$ between normalized returns $r^{norm}$ and normalized speeds $v^{norm}$ of drawdowns (red bars) and drawups (green bars, inverted x-axis) for different contracts and different probability levels $u$. The last row corresponds to the aggregated distributions. }
\label{fig:tail_dependence}
\end{figure}

On the contrary, the tail dependence $\lambda_u$ between returns and speed is 
significant and tends to increase for $u \to 1$, except very close to $1$ ($u>0.9997$)
due to the finite size of the data sample. The estimated value of 
the tail dependence $\lambda_u$ between returns and speed is approximatively 
in the range $0.09-0.12$ for drawdowns and $0.075-0.12$ for drawups, i.e., conditional
on a very large speed, there is about a 10\% probability that the corresponding
drawdown (drawup) is extreme in normalised return.
Figure~\ref{fig:tail_dependence}, which presents the tail dependence coefficients $\lambda_u$
at three probability levels $u=0.990, 0.995$ and $0.999$
for individual contracts, supports our previous findings at the aggregate level. With 
the exception of the OMXS, HCEI and ASX contracts that are characterised by monotonously decaying $\lambda_u$, all other analyzed future contracts exhibit clear signatures of non-zero tail dependence with $\lambda_u$ varying in the range $0.03-0.035$ (for NIKKEI, TOPIX) and $0.1-0.12$ (for CAC, DAX, AEX, STOXX, DJ, NIFTY). 

%===============================================================================
\section{Conclusion}\label{sec:conclusion}

We have investigated the distributions of $\epsilon$-drawdowns and $\epsilon$-drawups of 
the most liquid futures financial contracts of the world at time scales of $30$ seconds.
The $\epsilon$-drawdowns and $\epsilon$-drawups defined by expressions 
(\ref{eq:delta}) with (\ref{eq:cond}) are proposed as robust measures of the risks
to which investors are arguably the most concerned with. The time scale of $30$ seconds
for the time steps used to defined the drawdown and drawups is chosen as a 
compromise between robustness with respect to microstructure effects and 
reactivity to regime changes in the time dynamics. 
Similarly to the distribution of returns, we find that the distributions of 
$\epsilon$-drawdowns and $\epsilon$-drawups exhibit power law tails, albeit
with exponents significantly larger than those for the return distribution.
This paradoxical result can be attributed to (i) the existence of significant
transient dependence between returns and (ii) the presence of large outliers
(termed dragon-kings \citep{Sornette2009,SorouillonDK12}) characterizing
the extreme tail of the drawdown/drawup distributions deviating from the power law.
We present the generalised non-parametric DK-test together with a novel implementation of the 
parametric U-test for the diagnostic of the dragon-kings. Studying both the distributions
of $\epsilon$-drawdowns and $\epsilon$-drawups of
individual future contracts and of their aggregation confirm the robustness and
generality of our results. The study of the tail dependence between drawdown/drawup
sizes, speeds and durations indicates a clear relationship between size and speed 
but none between size and duration. This implies that the most extreme 
drawdown/drawup tend to occur fast and are dominated by a few very large returns.
These insights generalise and extend previous studies on outliers of drawdown/drawup
performed at the daily scale \citep{JohansenSornette2001JofRisk,JohSorepsidd10}.

%===============================================================================
\section*{Acknowledgments}

We would like to thank Professor Fr\'ed\'eric Abergel and the Chair of Quantitative Finance of l'\'Ecole Centrale de Paris (\url{http://fiquant.mas.ecp.fr/liquidity-watch/}) for the access to the high-frequency data used in the present analysis. We are very grateful to Professor Yannick Malevergne for many fruitful discussions while preparing this manuscript.

The statistical analysis of the data was performed using open source software: Python~2.7 (\url{http://www.python.org}) and libraries: Pandas~\citep{McKinney_Pandas2012}, NumPy~(\url{http://www.numpy.org/}), SciPy~(\url{http://www.scipy.org/}), IPython~\citep{Perez_IPython2007} and Matplotlib~\citep{Hunter_Matplotlib2007}.

%===============================================================================
%\section*{References}
%\bibliographystyle{elsarticle-harv.bst}
%\bibliography{/Users/vladimir/Work/Papers2/allpapers.bib}
%\bibliography{article_drawdowns_rev_v1.bib}

\end{document}